\documentclass{IEEEoj}
\usepackage{latexsym}

\usepackage[cmyk]{xcolor}
\usepackage{graphicx}
\usepackage[caption=false]{subfig}
\usepackage{amsfonts,amssymb,amsmath}
\usepackage{hyperref}
\usepackage{mathtools}

\def\nb0{{\mathbf{0}}}
\def\nb1{{\mathbf{1}}}

\usepackage{textcomp}
\usepackage{gensymb}
\usepackage{siunitx}
\usepackage{colortbl}
\usepackage{comment}
\usepackage{tikz}
\usepackage{cite}
\usetikzlibrary{chains,arrows,calc,positioning}
\usepackage{multirow}
\usepackage{color,soul}
\usepackage[para]{threeparttable}
\usetikzlibrary{patterns}

\def\SNR    {{\mathsf{SNR}}}
\def\INR    {{\mathsf{INR}}}
\newcommand{\wh}[1]{\widehat{{#1}}}
\newcommand{\bsym}[1]{\boldsymbol{{#1}}}
\newcommand{\subsf}{\sf \scriptscriptstyle}
\newcommand{\Herm}{{\subsf H}}

\def\BibTeX{{\rm B\kern-.05em{\sc i\kern-.025em b}\kern-.08em
    T\kern-.1667em\lower.7ex\hbox{E}\kern-.125emX}}
\AtBeginDocument{\definecolor{ojcolor}{cmyk}{0.93,0.59,0.15,0.02}}
\def\OJlogo{\vspace{-4pt}\hskip-4pt\includegraphics[height=18pt]{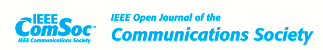}}

\begin{document}
\receiveddate{30 September, 2023}
\reviseddate{11 December, 2023 and 30 January, 2024}
\accepteddate{21 February, 2024}
\publisheddate{28 February, 2024}
\currentdate{27 February, 2024}

\doiinfo{OJCOMS.2024.011100}
\title{Cellular Wireless Networks \\ in the Upper Mid-Band}

\author{
        Seongjoon Kang\IEEEauthorrefmark{1} \IEEEmembership{(Graduate Student Member, IEEE)}, 
         Marco Mezzavilla\IEEEauthorrefmark{1} \IEEEmembership{(Senior Member, IEEE)}, 
         Sundeep Rangan\IEEEauthorrefmark{1} \IEEEmembership{(Fellow, IEEE)},
         Arjuna Madanayake\IEEEauthorrefmark{2} \IEEEmembership{(Member, IEEE)},
         Satheesh Bojja Venkatakrishnan\IEEEauthorrefmark{2} \IEEEmembership{(Senior Member, IEEE)}, 
        Gr{\'e}gory Hellbourg\IEEEauthorrefmark{3},
        Monisha Ghosh\IEEEauthorrefmark{4} \IEEEmembership{(Fellow, IEEE)},
        Hamed Rahmani\IEEEauthorrefmark{1} \IEEEmembership{(Member, IEEE)},
        Aditya Dhananjay\IEEEauthorrefmark{1}
        }
\affil{Electrical and Computer Engineering Department, New York University Tandon School of Engineering, Brooklyn, NY 11201, USA}

\affil{Department of Electrical and Computer Engineering, Florida International University, Miami, FL 33174 USA}

\affil{Department of Astronomy, California Institute of Technology, Pasadena, CA 91125, USA}

\affil{Department of Electrical Engineering, University of Notre Dame, Notre Dame, IN 46556, USA}

\corresp{CORRESPONDING AUTHOR: Seongjoon Kang (e-mail: sk8053@nyu.edu).}

\authornote{This work was supported in part by NSF under Grants 1952180, 2133662, 2236097, 2148293, 1925079, 2052764, 2229471, 2329012, 2216332, 1854798, 1902283, 1711625, 1509754, 1904382, 2243346, 2226392,  2128628, 2132700, 2229387, 2229428, 2229497 and 2128497; in part by the industrial affiliates of NYU Wireless  }

\begin{abstract}
The upper mid-band -- approximately from 7 to 24\,\si{GHz} --- has recently attracted considerable interest for new cellular services.  This frequency range has vastly more spectrum than the highly congested bands below \SI{7}{GHz} while offering more favorable propagation and coverage than the millimeter wave (mmWave) frequencies. 
In this regard, 
the upper mid-band
has the potential to provide a powerful and complementary frequency range that balances
coverage and capacity.
Realizing cellular networks that exploit the 
full range of these bands, however, presents significant technical challenges. 
Most importantly, spectrum will likely need to be shared with incumbents including communication satellites, military RADAR, and radio astronomy.   Also, the upper mid-band is simply a vast frequency range. Due to this wide bandwidth, combined with the directional nature of transmission and intermittent occupancy of incumbents, cellular systems will likely need to be agile to sense and intelligently use large spatial and frequency degrees of freedom.
This paper attempts to provide an initial assessment of the feasibility and potential gains of such adaptive 
wideband cellular systems operating across the upper mid-band. 
The study includes: (1) a detailed ray tracing simulation to assess potential gains of  multi-band systems in a representative dense urban environment and illustrate the value
of a wideband system with dynamic frequency selectivity;  (2) an evaluation of potential  cross-interference between satellites and terrestrial cellular services and interference nulling to reduce
that interference; and (3) design and evaluation of a compact multi-band antenna array structure.  Leveraging these preliminary results, we identify potential future research directions to realize next-generation systems in these frequencies. 

\end{abstract}

\begin{IEEEkeywords}
Upper mid-band, 6G, cellular wireless systems, FR3, satellite communications
\end{IEEEkeywords}

\maketitle

\section{Introduction}

Cellular wireless systems up to the fourth generation (4G)
had largely operated in a range of microwave frequencies below
\SI{6}{GHz}.  
Given the severe spectral shortage in these bands,
5G systems \cite{dahlman20205g,3GPP38300} 
introduced new capabilities in the millimeter
wave (mmWave) frequencies
above \SI{24}{GHz}
\cite{RanRapE:14,AkdenizCapacity:14,shafi20175g}.
The wide bandwidths available in the mmWave range, 
combined with spatial multiplexing capabilities, have 
enabled massive multi-Gbps
peak rates.  However, extensive measurements also now show that practical performance is often
intermittent
\cite{narayanan2021variegated,wei20225gperf}
with limited penetration inside \cite{narayanan2022comparative,rochman2022comparison}.  
At root, this poor coverage is
due to the inherently limited range of mmWave signals and their high susceptibility to blockage
\cite{Rappaport2014-mmwbook,slezak2018empirical,slezak2018understanding,maccartney2017rapid,zekri2020analysis}.

Against this backdrop, the upper mid-band spectrum, roughly from 7-24\,\si{GHz}, 
is being considered to provide a
good balance of coverage and bandwidth, overcoming the spectral
shortage of the sub-6 bands while having favorable propagation
and penetration relative to the mmWave frequencies. For these reasons, the band
has recently attracted
considerable attention for commercial
cellular communications and has been
cited by industry as a leading candidate spectrum for next generation
(NextG, 5G and beyond) wireless networks \cite{smee2022ten,samsung20226gspectrum,5Gamerica20223gpproadmap, lee2022extreme}.
The Federal Communications Commission (FCC) Technical Advisory Committee (TAC)
has recently identified the upper mid-band
as vital for meeting the growing data rate demands \cite{fcc2023preliminary,fcc20236Gworkinggroup}.  The 3rd Generation Partnership Project (3GPP), the organization that sets cellular standards, has also begun study of these bands \cite{3gpp38820} -- see Section~\ref{sec:background} for more details.

In 3GPP, the upper mid-band is referred to as
Frequency Range 3 (\textbf{FR3}).  
FR1 initially
referred to the traditional spectrum below \SI{6}{GHz}, which was later expanded to \SI{7.125}{GHz} \cite{sathya2020standardization}. 
FR2 refers to the mmWave spectrum that was introduced
in the fifth generation (5G) standard.

\begin{figure}  
    \centering
    \includegraphics[width=\columnwidth]{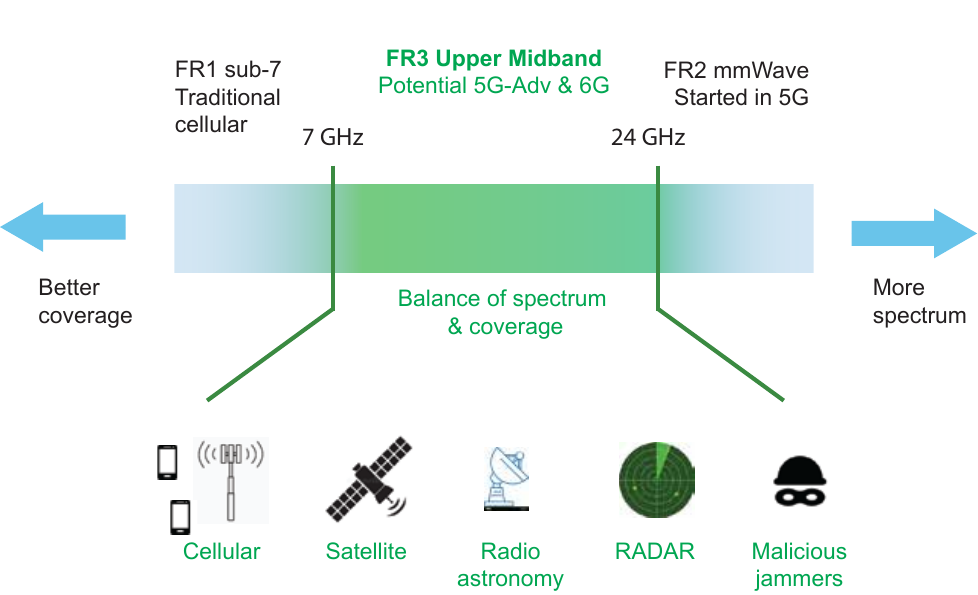}
    \caption{The \textbf{upper mid-band} (called frequency range 3 or \textbf{FR3} in 3GPP) is a potential 
    new band for the cellular services
    that offers a balance of coverage and spectrum.  
    To utilize the full band, cellular services will likely require
    sharing spectrum between satellite, 
    and radio astronomy along with resliency to 
    malicious jammers. }
    \label{fig:concept}
\end{figure}

\subsection{Challenges}
Development of cellular services in the full range of these bands faces significant challenges:
\begin{itemize}
\item \emph{Need for sensing and spectrum sharing with incumbents:}  Most importantly, these bands
are already used by several vital incumbent services. 
{This situation stands in contrast to the mmWave deployments
where the bands were significantly less occupied.}
As we review in Section~\ref{sec:incumbent_overview}, 
commercial satellite services already widely use these bands
and have been interested in
increased bandwidth allocations, particularly for
broadband rural access.  The upper mid-band is also vital for military and 
commercial RADAR and contains
scientifically important spectral lines that are fundamental 
for radio astronomy.  

\item \emph{Large spectral range:}  
The upper mid-band is simply a vast range of spectrum 
-- well beyond the span of most commercial cellular
front-ends that typically operate over a small percentage of
the carrier frequency.  This large range, combined
with the need to dynamically share spectrum
and directionally transmit, demands
new front-ends that adaptively and intelligently
sense and exploit large bandwidth and spatial
degrees of freedom.

\item \emph{Spectral resiliency: } 
Sensing capabilities, combined with the ability to 
adapt in space and frequency, 
can also yield improved defenses to external attacks from 
jammers and signal disruption.  
Such features can provide robust protection for
vital future cellular infrastructure.
\end{itemize}


\subsection{Contributions}
The broad goal of this paper is to assess the feasibility
and potential of agile wideband cellular systems operating
in these bands.  We will also identify
key challenges and areas for research. 
 Towards this end, our contributions are:
\begin{itemize}

    \item \emph{Channel modeling and 
    estimation of multi-band
    coverage}:  We briefly review 
    channel measurements in the upper mid-band range and discuss open areas of research
    (Section~\ref{sec:chanmeas}). 
    Also, based on current propagation models, we conduct a 
    ray-tracing simulation in 
    a representative dense urban area
    (New York City) to assess the coverage
    and capacity of a potential cellular 
    system in the upper mid-band (Section~\ref{sec:multibandcap}). Our
    results demonstrate that mobiles
    that can adaptively select one of multiple bands across the upper mid-band can improve coverage and data rates over mobiles
    restricted to a single band.  This
    finding argues for multi-band,
    adaptive systems to gain the full
    benefits of the frequency range.

    \item \emph{Interference with incumbents:}
    We perform an additional ray
    tracing simulation (Section~\ref{sec:sat_int}) to assess the
    potential interference of terrestrial cellular services on the satellite uplink.
    The analysis shows that interference
    from both user equipment (UEs) and base stations (BSs) can 
    be significant, but directional nulling
    can be effective in mitigating the 
    interference with some loss in the terrestrial network capacity.
    We also discuss some potential interference issues
    with passive radio astronomy in the 
    upper mid-band.

    \item \emph{Wideband antenna systems:}
    Finally, we present (Section~\ref{sec:antenna}) a compact,
    multi-band antenna array to demonstrate
    the feasibility of front-ends
    that operate over the entire upper mid-band in a small form factor.
    
\end{itemize}

\section{Background and Standardization Landscape}
\label{sec:background}


\subsection{Prior Developments in the Sub-6 and mmWave Bands}

The interest in the upper mid-band
has to be seen in the context of 
developments of commercial cellular systems in both the sub-7 and mmWave
frequency ranges over the last decade.
Early mmWave experiments and capacity analyses
such as
\cite{akoum2012coverage,rappaportmillimeter,RanRapE:14,AkdenizCapacity:14}
suggested the possibility of massive data rates
in micro-cellular deployments owing to 
the wide bandwidths and spatial multiplexing gains available in the mmWave frequencies.  
Based partially on these and other results, 
the mmWave bands emerged as an integral component of the 5G New Radio (NR) specification \cite{dahlman20205g,3GPP38300}.  Commercial mmWave deployments appeared shortly after the release of the specification,
particularly in the US \cite{shafi20175g}.

Coverage at reasonable cell densities
in these deployments, however, has been an enormous challenge.  
Several recent measurements in commercial mmWave 
networks now clearly demonstrate that outdoor coverage can be highly intermittent 
\cite{narayanan2021variegated,wei20225gperf}
with limited penetration indoors  \cite{narayanan2022comparative,rochman2022comparison}.
MmWave signals are simply blocked by many common building materials such as concrete \cite{Rappaport2014-mmwbook,zekri2020analysis} 
as well as the human body and other obstacles \cite{slezak2018empirical,slezak2018understanding,maccartney2017rapid}.
See also Section~\ref{sec:o2i} below.

Parallel to the developments in the mmWave,
significant spectrum was also released in portions of the mid-band
frequencies above \SI{3}{GHz}, including the  Citizens Broadband Radio Service
(CBRS) band (3.55 to 3.7\,\si{GHz}) \cite{fcccbrs}
and the C-band
\cite{FCC20-22} (3.7 to 4.2\,\si{GHz}).
These bands were instrumental for both so-called private
5G networks 
\cite{aijaz2020private} as well as spectrum expansion 
in many wide area public networks.
More recently, there has also been significant discussion of unlicensed bands from approximately 6 to 7\,\si{GHz},
extending the mid-band further
\cite{sathya2020standardization}.
In particular, both wireless LAN
and cellular services have been considering enhancements to operate in these bands \cite{IEEE80211ax,naik2021coexistence}.

Measurements from commercial network
monitoring companies, now offer an opportunity to compare the
practical performance of these networks.
On the one hand, Ookla reported in 2022 \cite{ookla2023mmwaverates} that 
the tested 
mmWave networks offer an incredible \SI{1.6}{Gbps} median downlink throughput, up
to seven times higher than systems in C-band.
Similarly, in 2021, Open Signal
demonstrated \cite{opensignal2021mmwave}  that Verizon's mmWave downlink capacity
was at least three times as high as networks
relying mostly on mid-band spectrum.
Nevertheless, networks  with
mid-band alone often provided downlink
speeds in the hundreds of Mbps.  Moreover,
the coverage for mid-band networks was much more uniform
than mmWave.  
For example, Open Signal's study
\cite{opensignal2021mmwave} showed that users
connect to mmWave less than 1\% of the time.

\subsection{Recent Standardization Efforts in the Upper Mid-band}
The relative success of commercial cellular systems in the mid-band, combined with the high data rates, but intermittent coverage, of mmWave has set the stage for
interest in the upper mid-band.
The hope is to provide the high data rates
close to those available in the mmWave range,
but with much more uniform coverage.
As mentioned in the Introduction, the bands have been identified by industry \cite{smee2022ten,samsung20226gspectrum}
and 
3GPP has recently formally started study of services in the 7 to 24\,\si{GHz} band \cite{3gpp38820,5Gamerica20223gpproadmap},
where they are called frequency range 3 or FR3.
Note that, in 3GPP terminology, frequency range 1, or FR1, has been extended from sub-6 GHz to sub-7 GHz to include the frequencies in the 6 to 7\,\si{GHz} range
in FR1~\cite{3GPP38921}.  

In 2022, the FCC also began discussion on two \SI{500}{MHz} 
bands from 12.2-12.7\,\si{GHz} \cite{FCC20-443},
and 12.7-13.2\,\si{GHz}\cite{FCC22-352} for possible cellular use.  While the first
band has now been rejected for cellular services, there is wide recognition of the
need for opening the upper mid-band
for cellular services.  For example,
the FCC Technical Advisory Committee (TAC)
has recently published a comprehensive survey of spectrum from 7.125 to 24\,\si{GHz} 
\cite{fcc2023preliminary}
as well as a 6G working paper \cite{fcc20236Gworkinggroup}.
These analyses emphasize the need for
considering larger portions of the upper mid-band for terrestrial cellular services
in order to meet the growing data demand.

\subsection{Incumbency and the Need for Spectrum Sharing}
\label{sec:incumbent_overview}
The FCC TAC analyses \cite{fcc2023preliminary,fcc20236Gworkinggroup}
also emphasize that 
a key issue in allocating the upper mid-band
is incumbents, particularly commercial satellite services that also need bandwidth.
In fact, the rejection of the 
12.2-12.7\,\si{GHz} band proposed in \cite{FCC20-443} for cellular services 
was largely due to 
the interference onto ground satellite
units.   We will 
perform some simple cellular-satellite 
interference calculations in Section~\ref{sec:sat_int}.

The FCC analysis \cite{fcc2023preliminary}
has thus considered several
potential models for spectrum sharing
in the upper mid-band.  
Interestingly, spectrum sharing was critical
in CBRS and C-Band allocations \cite{mueck2015spectrum,matinmikko2020spectrum}.
Spectrum sharing mechanisms in these
bands included spectrum access systems
SAS and more general licensed spectrum access (LSA) schemes are now being considered in the upper mid-band as well.
More dynamic methods for spectrum sharing,
as is being considered by 3GPP 
\cite{3GPP21917,QcomDSS} could also be used.

In short, the upper mid-band is a vast
and valuable frequency range for numerous
services.  
How to allocate and share the spectrum 
between various users will be one of the
fundamental design and policy challenges
going forward.

\begin{figure*}[t]
  \centering
  \subfloat[\textcolor{black}{3D model}]{
    \includegraphics[trim={3cm 1cm 0 22cm},clip,width=0.4\textwidth]{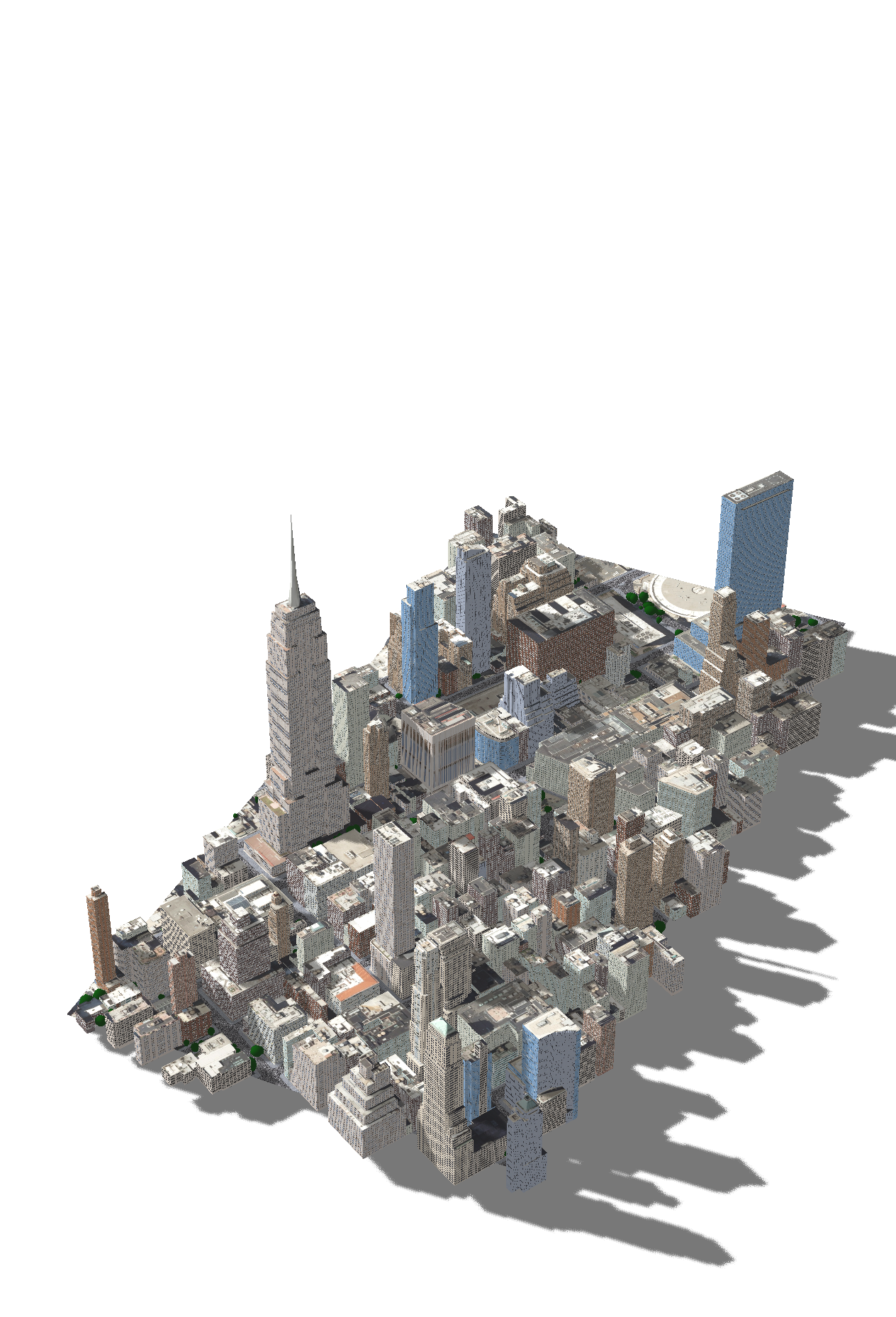}
    \label{fig:nyc_map}
  }
  \hfill
  \subfloat[gNodeB arrays]{
    \includegraphics[trim={0 0 0 3mm}, width=0.22\textwidth]{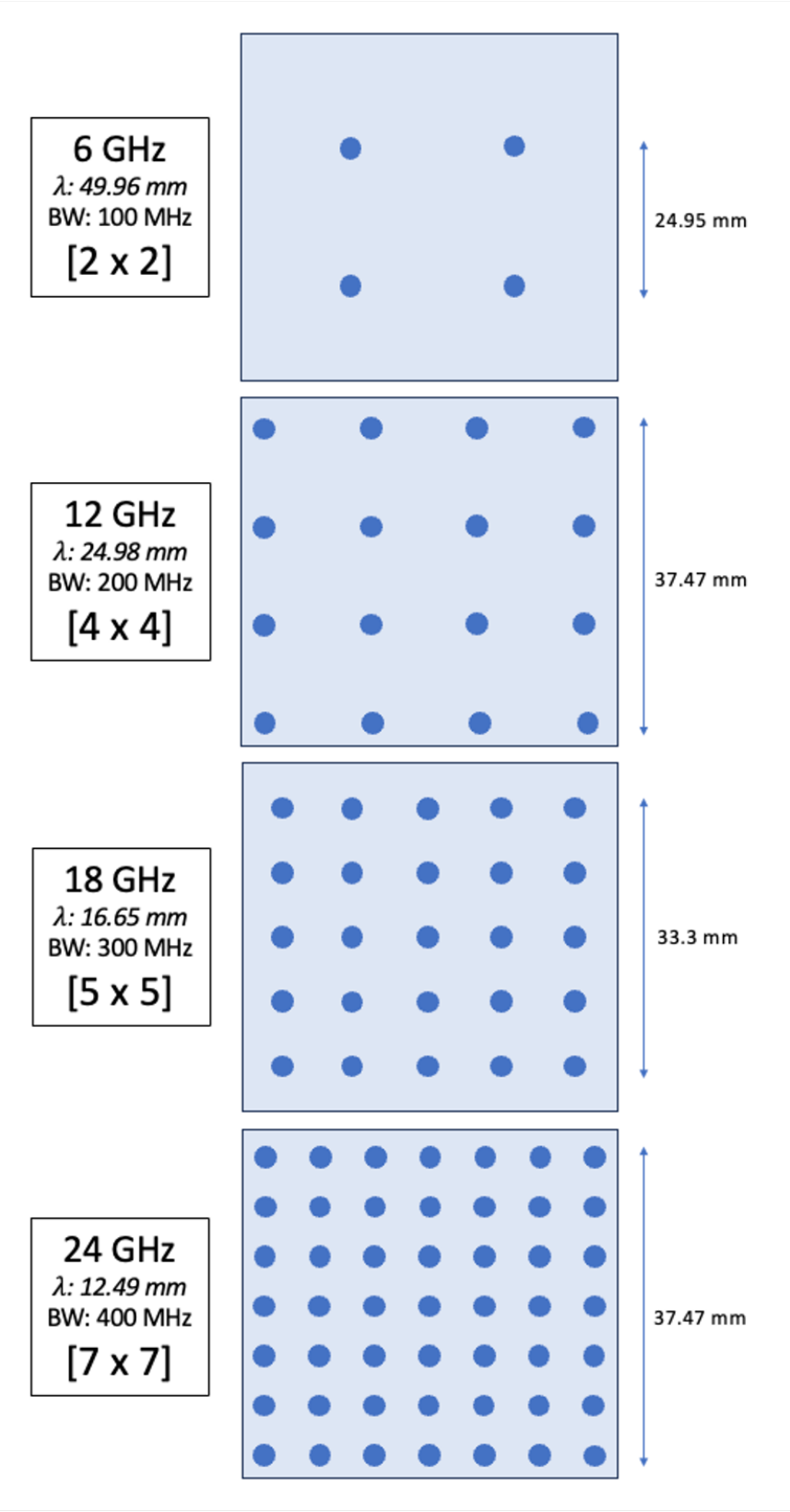}
    \label{fig:array_arch}
  }
  \hfill
  \subfloat[\textcolor{black}{Coverage}]{
    \includegraphics[width=0.3\textwidth, height = 7.3cm]{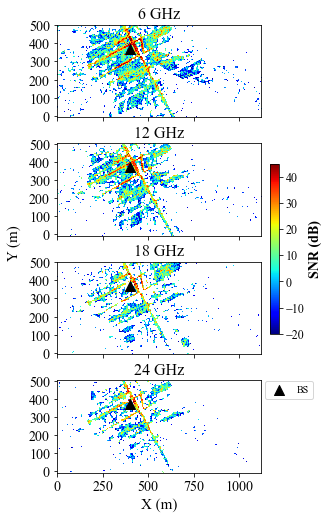}
    \label{fig:coverage_plot}
  }
  \caption{Simulation results obtained via ray-tracing a dense urban area of NYC at four frequencies in the upper mid-band
  for one example BS site.  Left:  3D model of the NYC area (Herald Square) that was used for raytracing. The model was downloaded from~\cite{geopipe}, which enabled a fairly accurate mapping of foliage and building materials.  Center:  gNodeB antenna array architecture and bandwidth at each carrier frequency.
  Right:  Coverage map obtained from raytracing data from an example single BS.  The reduced coverage with higher frequencies is readily visible.}
  \label{fig:simulations}
\end{figure*}
\begin{figure}[ht]
    \centering
    \includegraphics[width=0.9\columnwidth]{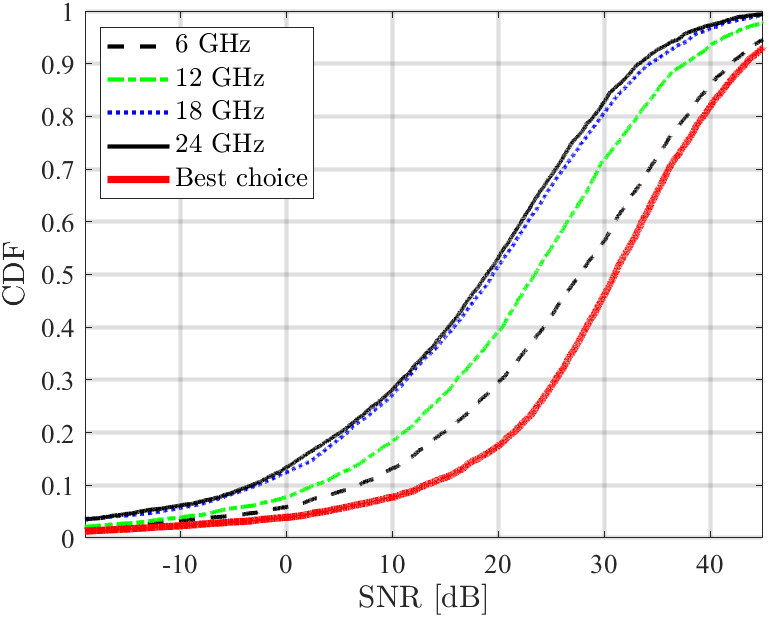}
    \caption{Aggregate multi-band SNR distribution for all \textbf{outdoor} UEs
    \textbf{without blockage}.}
    \label{fig:snr_plot_outdoor}
\end{figure}
\begin{figure*}[ht]
  \centering
  \subfloat[Data rate CDF \textbf{without blockages}]{
    \includegraphics[width=0.45\textwidth]{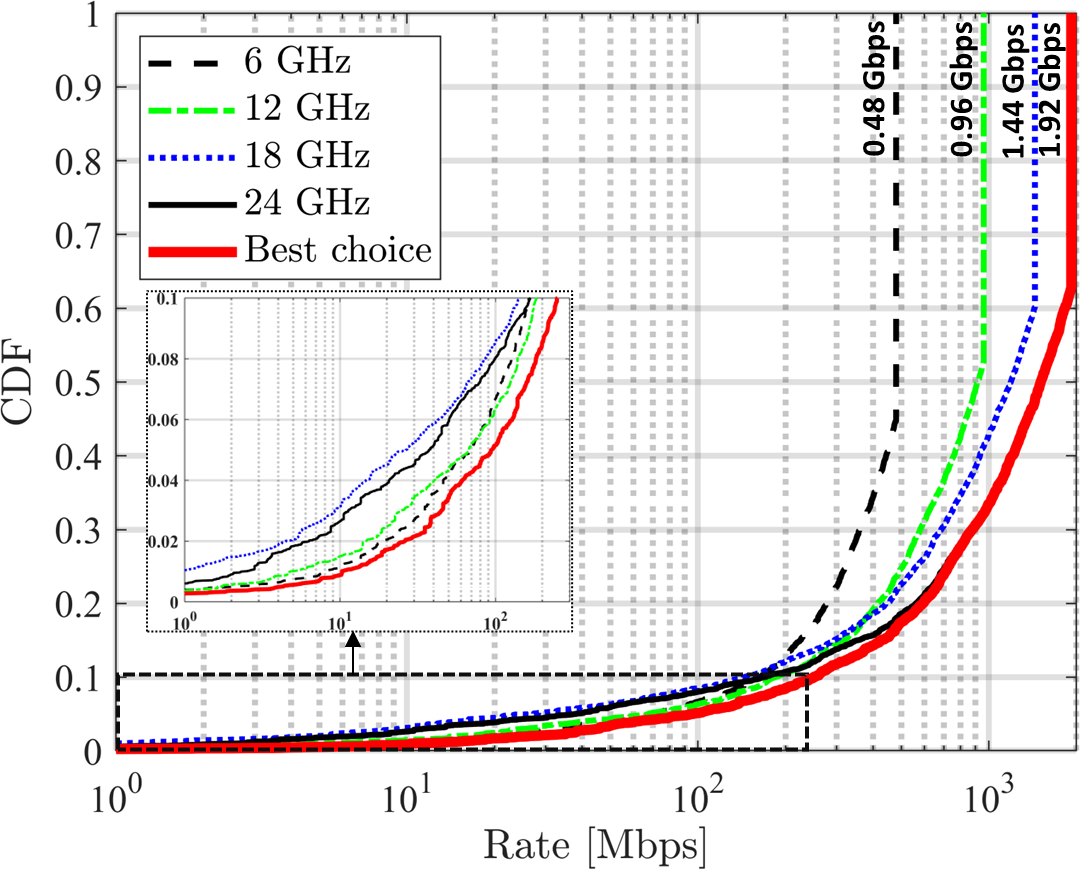}
    \label{fig:rate_plot_outdoor}
  }
  \hfill
  \subfloat[Data rate CDF \textbf{with blockages}]{
    \includegraphics[width=0.45\textwidth]{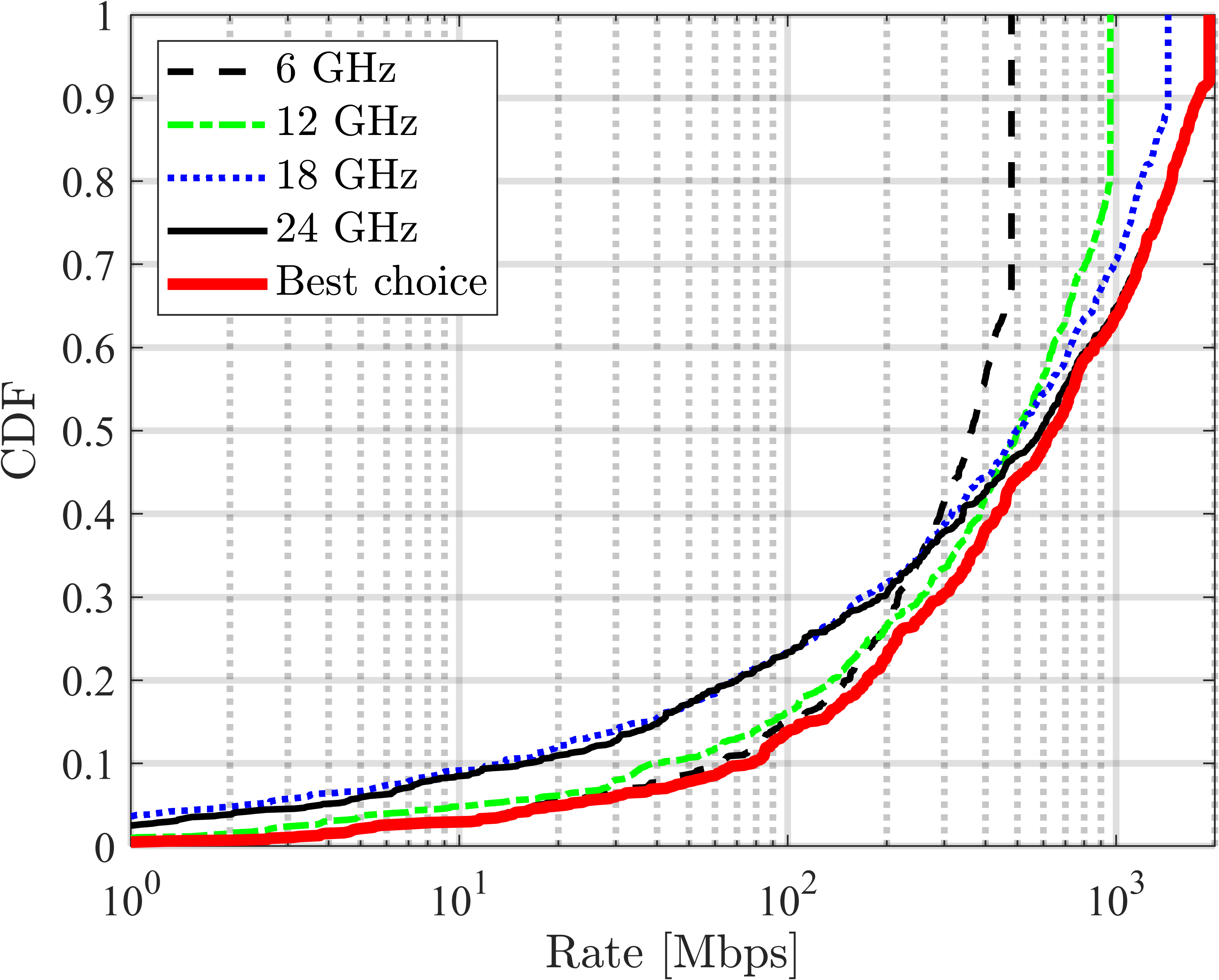}
    \label{fig:rate_plot_outdoor_with_blockage}
  }
  \caption{Aggregate multi-band rate distribution for all \textbf{outdoor} UEs.}
    \label{fig:snr_rate_plot_outdoor}
    
  \end{figure*}

\section{Coverage and Capacity Gains}

\subsection{Prior Channel Measurements and Studies} 
\label{sec:chanmeas}
Aspects of channel propagation in the upper mid-band have 
been studied for over two decades.  
For example, \cite{janssen1992propagation,janssen1996wideband} studied performed indoor radio measurements up to \SI{11.5}{GHz} in the early and mid-1990s. Significant research has continued.
For example, additional indoor measurements can be found in \cite{de20182} (from 2.4 to 61,\si{GHz}) and in corridors
from 9 to 11,\si{GHz} in \cite{batalha2019indoor}.
Attenuation measurements for various building materials at 800 MHz to 18\,\si{GHz}
were conducted in \cite{deng2016indoor}.
Indoor and outdoor 5G diffraction measurements at 10, 20 and 26\,\si{GHz} can be found in
\cite{rodriguez2014radio}.  The work in \cite{kristem2018outdoor} performed wideband outdoor channel measurements at 3--18\,\si{GHz}
and a more recent comprehensive set of outdoor and indoor measurements
at 3.3, 6.5, 15 and 28\,\si{GHz} was conducted in \cite{miao2023sub}.
Satellite propagation has also been extensively studied -- see, for example, \cite{pan2001high} on earth satellite measurements at \SI{12}{GHz}.

A key and consistent
finding of these channel measurement studies is that the factors that influence large-scale propagation,
such as transmission losses, reflectivity, and diffraction, vary considerably over the upper mid-band.
As expected, the lower frequencies in the upper mid-band provide more favorable coverage in most
indoor and outdoor scenarios.  At the same time, higher frequencies offer better bandwidth
since bandwidth allocations generally scale with the carrier frequency. 

\subsection{Multi-frequency Outdoor Capacity Gains}
\label{sec:multibandcap}
One implication of the variability
of propagation across the upper mid-band is that cellular systems would ideally have access to bands across the spectrum with real-time band selection.
Wideband mobiles could use the higher frequencies with higher bandwidth when
coverage is available, and switch
to lower frequencies when the higher frequencies are blocked. 

\begin{table}[t]
\footnotesize
  \begin{center}
    \caption{Multi-frequency capacity simulation parameters.}
    \label{tab:cap_sim}
\begin{tabular}{|>{\raggedright}p{3cm}|>{\raggedright}p{4.5cm}|}
\hline
\cellcolor{purple!15} \textbf{Parameters} & \cellcolor{purple!15} \textbf{Values} 
\tabularnewline \hline 
      {Area (m$^2$) } & $1120 \times 510$
     \tabularnewline \hline
      {Inter-site distance between gNBs (m)} & $200$
      \tabularnewline \hline
         {Maximum transmit power of gNBs (\text{dBm})} & \SI{33}{dBm} (from 3GPP TR 38.141~\cite{3GPP38141}, Table 6.3.1-1, Local Area BS) 
      \tabularnewline \hline
       Number of sectors for gNBs & $3$  
      \tabularnewline  \hline
       {Down-tilted antenna angle of gNBs} & $-12$\degree  
      \tabularnewline  \hline
      UE Noise figure (dB)  & 7 
      \tabularnewline  \hline
      {Field pattern of antenna element} & Taken from 3GPP TR 37.840 \cite{3GPP37840}
      \tabularnewline  \hline
    \end{tabular}

\begin{tabular}{|>{\raggedright}p{3cm}|>{\raggedright}p{0.8cm}|>{\raggedright}p{0.8cm}|>{\raggedright}p{0.8cm}|>{\raggedright}p{0.8cm}|}
\hline
\multicolumn{5}{|l|}{\cellcolor{purple!15} Frequency-dependent parameters}
\tabularnewline \hline
Frequency [GHz] & 6 & 12 & 18 & 24 \tabularnewline \hline
Bandwidth [MHz] & 100 & 200 & 300 & 400 \tabularnewline \hline
BS URA dimensions & $2 \times 2$ & $4 \times 4$ & $5 \times 5$ & $7 \times 7$   \tabularnewline \hline
UE ULA dimensions & $1 \times 2$ & $1 \times 2$ & $1 \times 3$ & $1 \times 3$ \tabularnewline \hline
\end{tabular}
  \end{center}
\end{table}

To estimate the potential gain of such a multi-band, frequency adaptive system in the
upper mid-band, we consider a dense urban area, i.e., Herald Square in New York City, as shown in Fig.~\ref{fig:nyc_map} and downlink scenario (gNB$\rightarrow$UE).
In dense urban scenarios, capacity a key
requirement.  At the same time, providing satisfactory coverage is challenging at high frequencies due to blockage, as have been experienced in the mmWave bands \cite{slezak2018empirical,slezak2018understanding,jain2018driven}.
Within this area, we consider a hypothetical
systems operating at up to four potential frequencies: 6, 12, 18, and 24\si{GHz}.

The simulation parameters of the potential cellular downlink systems at these frequencies are shown in Table~\ref{tab:cap_sim}. 
Note that the bandwidth in each frequency scales with the carrier, as is typical in deployments today.
As shown in Fig.~\ref{fig:array_arch}, the size of the gNB antenna array also scales so that the aperture is approximately constant.
Furthermore, in the simulation area, we manually selected the locations of 18 BSs (gNB) on rooftops, corresponding to an inter-site distance (ISD)
of approximately \SI{200}{m}, typical for urban microcellular evaluations.
Terrestrial UEs are randomly placed outside buildings, and ray tracing was performed using Wireless Insite~\cite{Remcom}, \textcolor{black}{which has recently been proven fairly accurate by conducting real-world measurements in the mmWave bands, as reported in~\cite{de2023ray, remcommpresnt2017modeling}}.

{As mentioned in the Introduction, a critical factor in the performance
of mmWave systems is the susceptibility of signals to blockage
\cite{Rappaport2014-mmwbook,slezak2018empirical,slezak2018understanding,maccartney2017rapid,zekri2020analysis}.
To model the effect of blockage, we consider two scenarios:
\begin{itemize}
    \item \textbf{No blockage}; and
    \item \textbf{Blockage} modeled with the 3GPP Blockage Model B \cite{3GPP38901} with $K=4$
    random human blockers.
\end{itemize}
Fig.~\ref{fig:coverage_plot} shows a coverage map of the unblocked wideband SNR 
for a single BS deployment as an example.  
As expected, we see that coverage is greatly reduced as frequency increases.}

For both the unblocked and blocked cases, 
we compute the wideband SNR for each UE and BS and assume the UE is served
by the BS with the strongest unblocked SNR.
Then, given an $\SNR$, we assume the achieved rate $R$ (goodput) follows a standard realistic model \cite{mogensen2007lte}:
\begin{equation} \label{eq:ratesnr}
    R =  B \min\{ \rho_{\rm max}, \alpha \log_2(1 + \SNR))
\end{equation}
where $B$ is the bandwidth, $\alpha$ is a system bandwidth loss factor to account for overhead
and receiver imperfections, and $\rho_{\rm max}$ is a maximum spectral efficiency. 
{Following~\cite{mogensen2007lte}, we adopt $0.57$ for $\alpha$ and $4.8$ for $\rho_{\rm max}$}.

Fig.~\ref{fig:snr_plot_outdoor} shows the resulting SNR distribution
for frequencies 6, 12, 18 and 24~\si{GHz} at all outdoor UE locations in the study
area without blockage.  As can be seen in Fig.~\ref{fig:snr_plot_outdoor}, UEs at lower frequencies experience uniformly better SNRs than higher frequencies due to
favorable propagation and reduced noise power from
using a smaller bandwidth compared to higher frequencies. 
Fig.~\ref{fig:rate_plot_outdoor} shows the corresponding
rate distribution.  
{We observe that, even though the SNRs are lower, as expected,
the use of higher frequencies (18 and 24~\si{GHz}) ensures superior data rate due to the wider bandwidth.
Note that from \eqref{eq:ratesnr}, there is a maximum rate of 
$R = B\rho_{\rm max}$ which corresponds to $0.48$, $0.96$, $1.44$ and $\SI{1.92}{Gbps}$
at frequencies $6$, $12$, $18$ and $24$~\si{GHz}, respectively.  
Hence, higher peak rates can be achieved at higher frequencies, where more bandwidth is available. 
However, UEs at the cell edge (for example, UEs up to the bottom 10\% percentile of the CDF) suffer a significantly worse rate in the higher frequency bands (18 and 24~\si{GHz}) compared to UEs at lower frequencies (6 and 12~\si{GHz}).  This property is also expected, as propagation is much less favorable for UEs at the cell edge at higher frequencies.}

{ Fig.~\ref{fig:rate_plot_outdoor_with_blockage} 
shows the rate distribution with blockage.  We observe that the cell edge rates at the high frequencies drop sharply.  Indeed, UEs in $18$ and $\SI{24}{GHz}$ show significantly worse performance approximately $35\%$ of the time, compared to UEs in $6$ and $\SI{12}{GHz}$. Thus, when blockers, such as humans or vehicles, surround an UE,
as would occur commonly in an urban scenario, the high-frequency coverage suffers significantly, while the low-frequency coverage is significantly more uniform.
}

The curves labeled ``best choice" in Fig.~\ref{fig:rate_plot_outdoor} and Fig.~\ref{fig:rate_plot_outdoor_with_blockage} correspond to the rates
for the UEs that select the best BS and frequency.  
As expected, the best choice is uniformly better than any individual frequency.
Indeed, the best choice achieves the same peak rates as the UEs in the high-frequency bands, along with the improved cell edge rates of the low frequencies.
These results motivate wideband cellular systems that adaptively select across a range of bands. 
 Such systems can obtain the bandwidth benefits at high frequencies while providing
robustness and resistance to blockage at low frequencies.

\begin{figure}[t!]
    \centering
    \includegraphics[width=0.9\columnwidth]
    {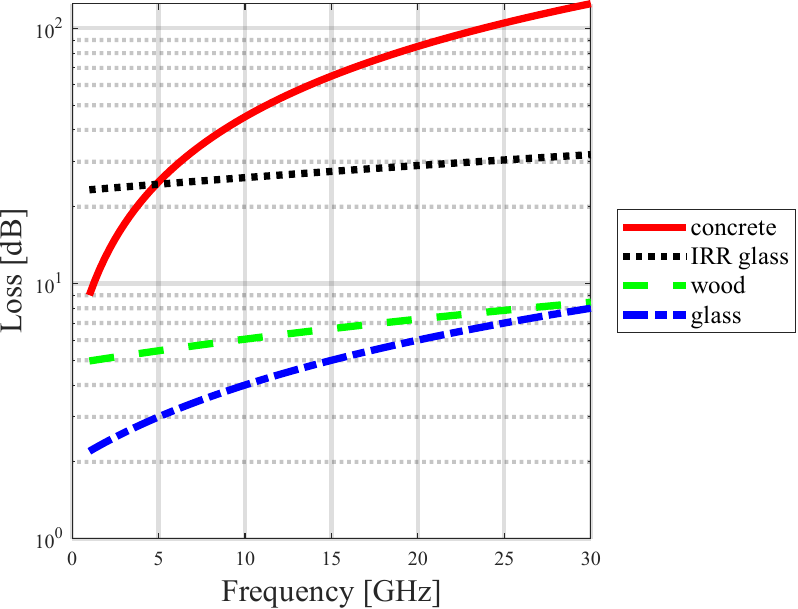}
    \caption{3GPP model \cite[Table 7.4.3-1]{3GPP38901} for outdoor-to-indoor penetration losses of various
    indoor materials at different frequencies.}
    \label{fig:o2i_loss}
\end{figure}

\subsection{Penetration Loss}
In the previous section, we focused on outdoor UEs. 
Coverage for indoor from outdoor cell sites is called 
outdoor to indoor (O2I) penetration.  To understand the potential for O2I coverage in the upper mid-band, Fig.~\ref{fig:o2i_loss}
plots the 3GPP model \cite[Table 7.4.3-1]{3GPP38901} for the loss of the O2I path for several common exterior building materials.  For each material,
the 3GPP model for pathloss is given by 
\begin{equation} \label{eq:3gpp_o2i}
    L = a + bf
\end{equation}
where $L$ is the path loss in dB, $f$ is the frequency
in GHz, and $a$ and $b$ are linear constants that
depend on the carrier frequency and are given in \cite[Table 7.4.3-1]{3GPP38901}.  Note that the curve in
Fig.~\ref{fig:o2i_loss} does not appear linear
since the pathloss is plotted on a logarithmic scale.
We see that standard glass and wood are relatively permeable throughout the frequency range ($<$\SI{8}{dB}
for both materials),
while infrared reflecting glass (IRR) is relatively impermeable across all RF frequencies ($>$\SI{20}{dB}).  
Concrete, however, makes a sharp transition
from relatively permeable to inpenetrable, precisely
in the upper mid-band.


\setlength{\belowcaptionskip}{-10pt}
\begin{figure}[t!]
  \centering
  \subfloat[SNR CDF]{
    \includegraphics[width=0.45\textwidth]{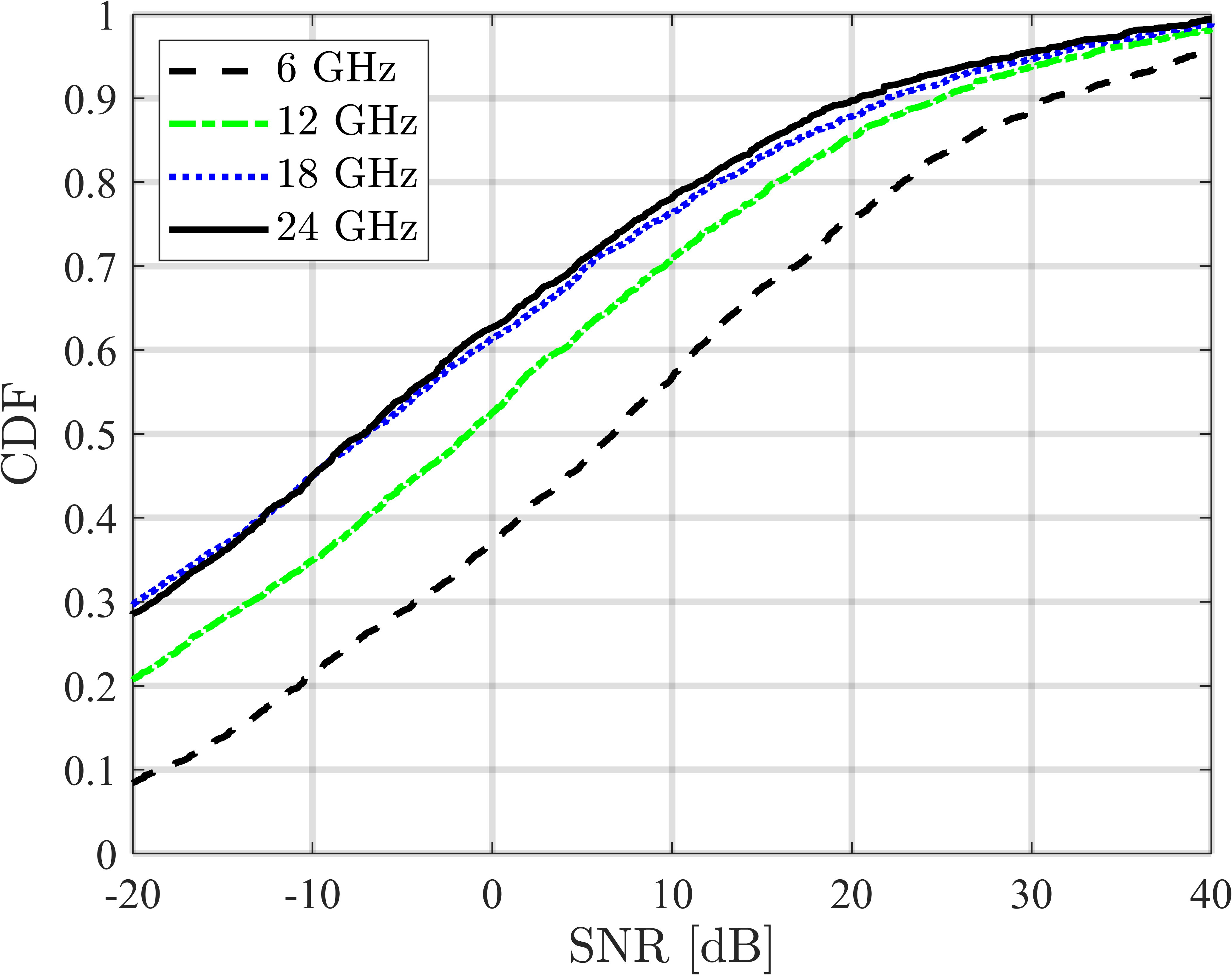}
    \label{fig:snr_plot_indoor}
  }
  \hfill
  \subfloat[Data rate CDF]{
    \includegraphics[width=0.45\textwidth]{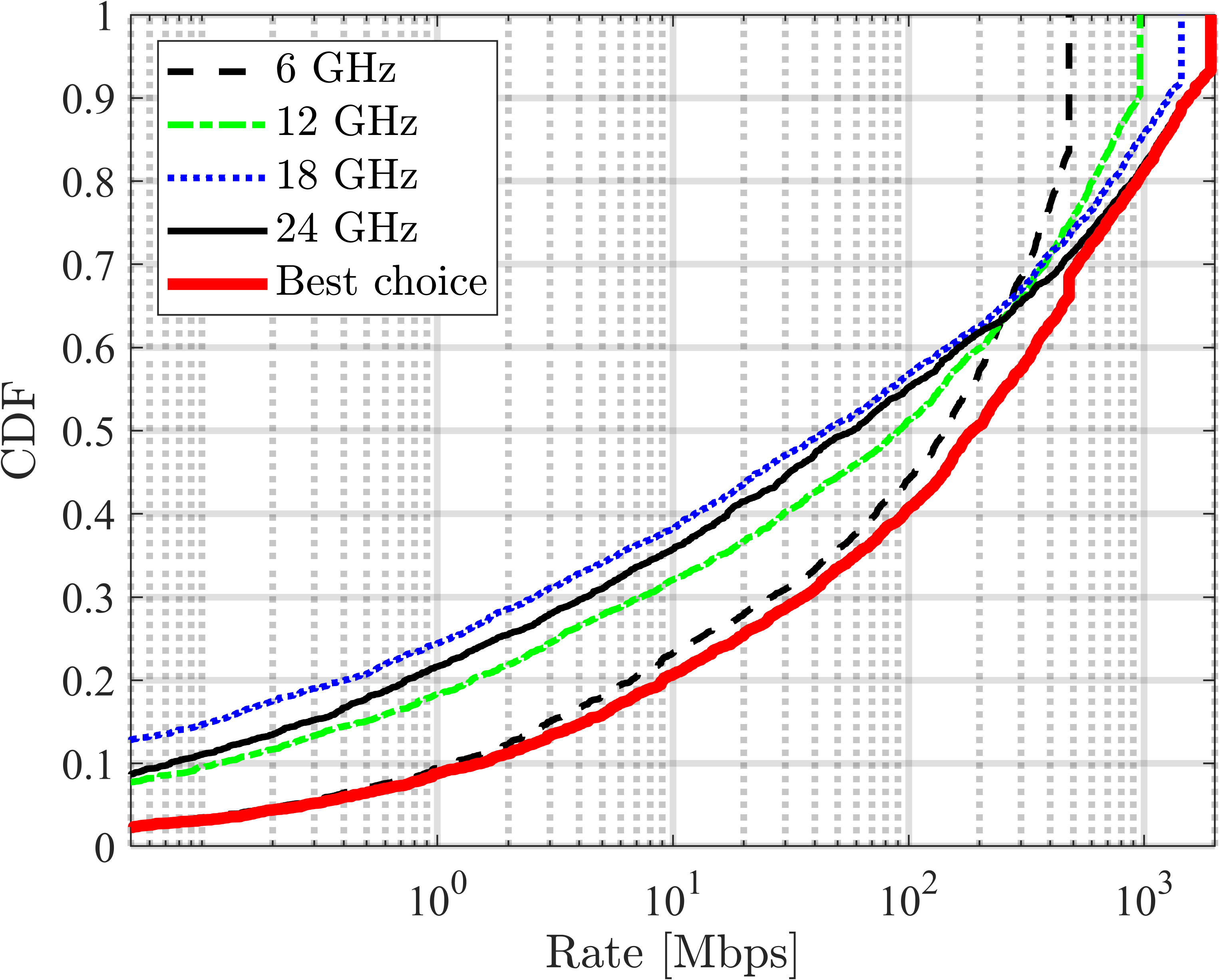}
    \label{fig:rate_plot_indoor}
  }
  \caption{Aggregate multi-band SNR and rate distribution for all \textbf{indoor} UEs.}
  \label{fig:snr_rate_plot_indoor}
\end{figure}
\setlength{\belowcaptionskip}{10pt}

\subsection{Multi-frequency Indoor Capacity Gains}
\label{sec:o2i}

To understand the effect of penetration loss on capacity, we re-run the simulation
with users randomly placed indoors.  Note that 
the 3D models used in the ray tracing
have approximate material classifications for
the facades of each building. 
Using the penetration loss models 
from Eq.~\ref{eq:3gpp_o2i} combined with the exterior wall classification, Fig.~\ref{fig:snr_rate_plot_indoor}  plots the estimated CDFs of the SNR and data rate for 
\emph{indoor} UEs.  Fig.~\ref{fig:snr_plot_indoor} shows that
the SNR for indoor UEs is considerably reduced at higher frequencies due to high penetration loss, as mentioned in the previous section. 
In this setting, concrete is a dominant
exterior wall material and significantly
reduces the signal penetration -- see Fig.~\ref{fig:o2i_loss}.
Indeed, Fig.~\ref{fig:rate_plot_indoor} shows that, for roughly 65\% of the users, the data rates at lower frequencies are higher than at higher frequencies. 
These results indicate a further
potential gain of wideband cellular
systems in FR3: the lower
frequencies can provide valuable
indoor coverage behind materials such
as concrete while the higher
frequencies can opportunistically
offer high capacity for outdoor users
and selected indoor users with minimal
blockage (e.g., indoor users next to non-IRR
glass windows).



\begin{figure}[t!]
\centering
\subfloat[SINR CDF]{
\includegraphics[width=0.45\textwidth]{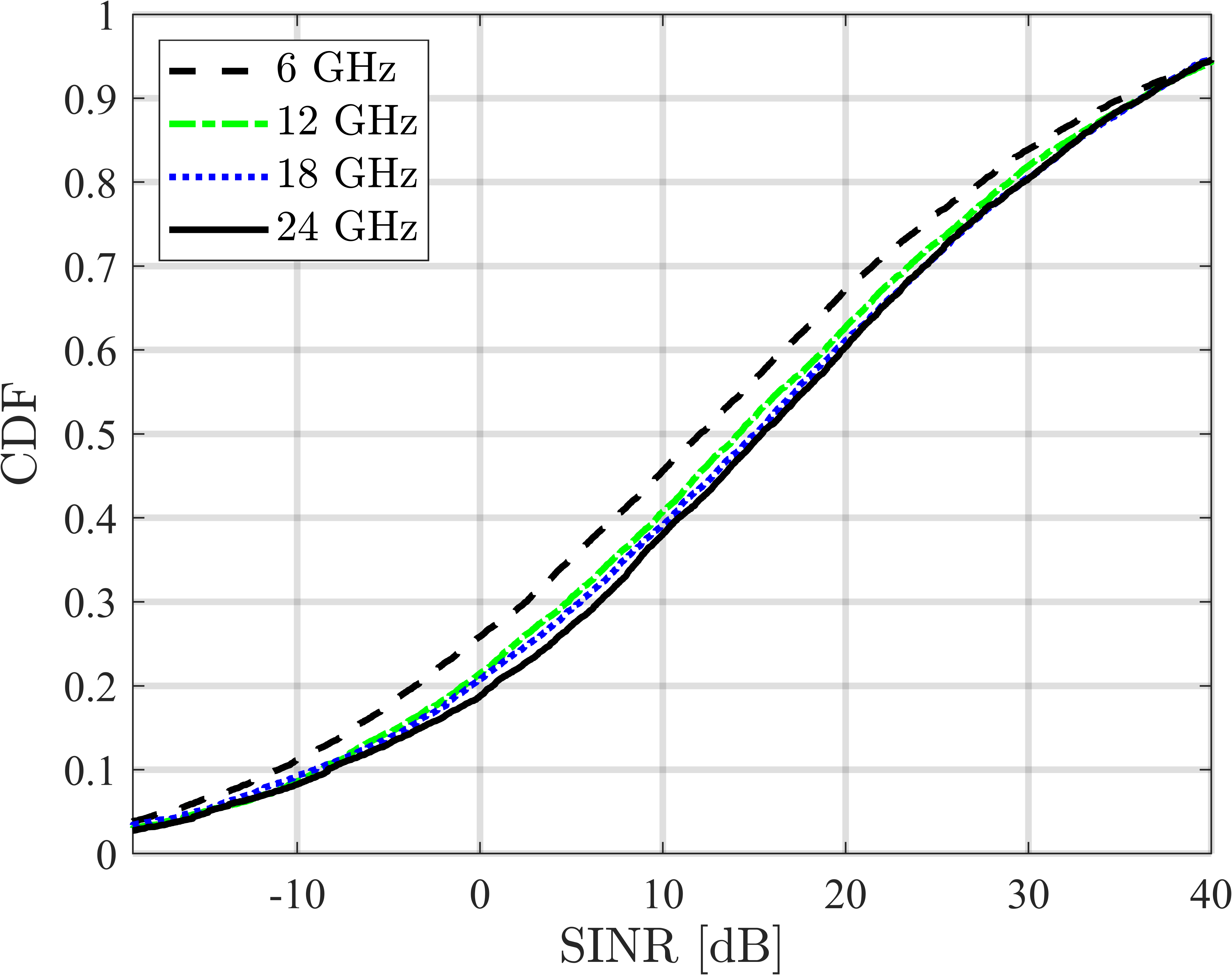}
\label{fig:sinr_plot}
}
\\
\subfloat[Data rate CDF]{
\includegraphics[width=0.45\textwidth]{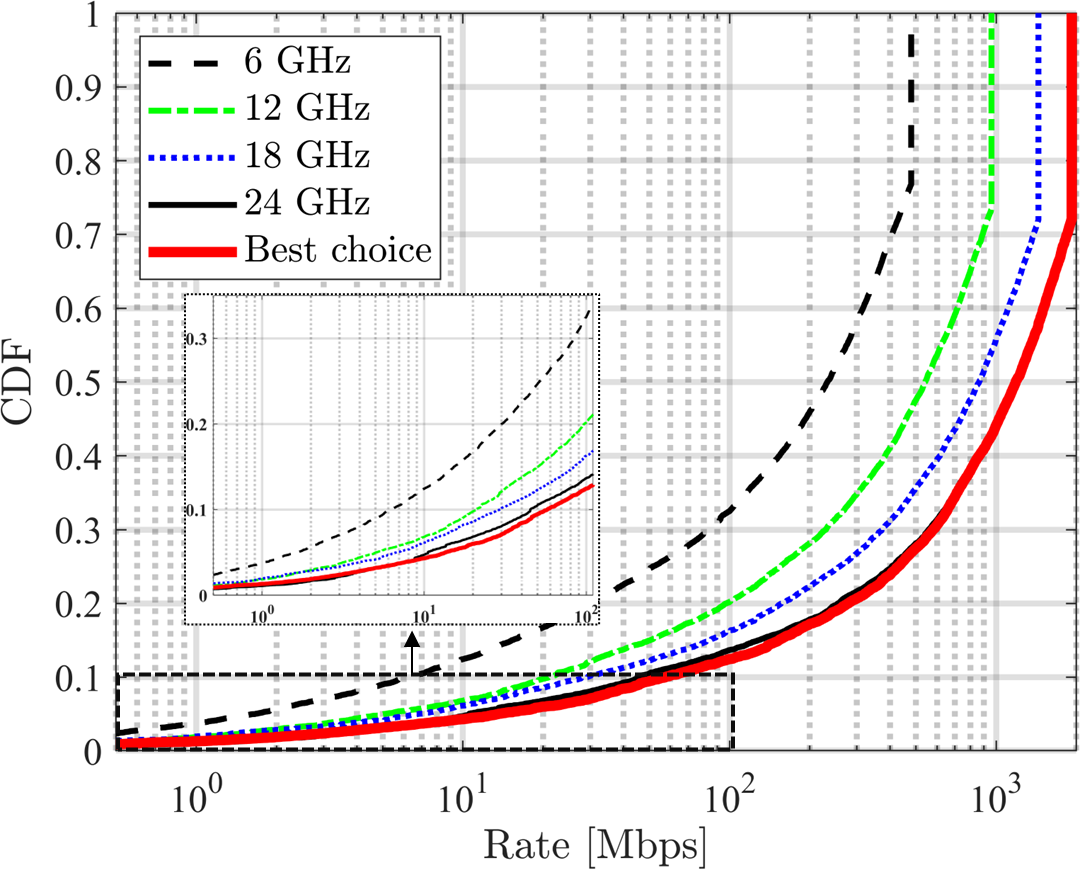} 
\label{fig:rate_plot_from_sinr}
}
\caption{Aggregate multi-band SINR and rate distribution for all \textbf{outdoor} UEs
in a multi-cell setting with interference.}
\label{fig:sinr_rate_plot}
\end{figure}


\subsection{Effects of Interference}
As the last discussion, we consider a full buffer interference scenario where BSs transmit data to their associated outdoor UEs simultaneously, and each UE receives interference signals from other BSs. 
 Note that this is the worst-case scenario to show the lower bound of the system capacity gains.
As before, considering $\SI{200}{m}$ ISD, $18$ BSs are manually selected and concurrently transmit data to $18$ UEs.

Fig.~\ref{fig:sinr_rate_plot} shows the CDF of the signal-to-interference-noise ratio (SINR) and the corresponding data rate calculated using \eqref{eq:ratesnr}. As shown in Fig.~\ref{fig:sinr_plot}, SINRs are degraded at lower frequencies (6 and \SI{12}{GHz}) because the small number of antenna arrays produces wider beamwidths. 
In contrast, due to directional transmissions with narrow beamwidths, 
UEs at higher frequencies experience less interference.

As a result, we observe in Fig.~\ref{fig:rate_plot_from_sinr} that data rates at higher frequency are almost uniformly better than those at lower frequency.  
Thus, the ``best choice"  always selects the high-frequency band.
{
This result suggests yet another benefit of adaptive systems: in interference-limited scenarios, 
BSs or UEs can automatically choose bands where directionality can reduce interference.
}

{
Note that to clearly see the effect of interference, we have not added blocking
in this simulation as we did earlier.  If blocking were added,
then the high frequency performance would degrade and not be uniformly better.  The point is that adaptive systems can naturally select the optimal bands with multiple factors, including
path loss, penetration, and directionality.
}

\section{Interference with Incumbent Services}
\subsection{Commercial Communication Satellites}
One of the most vital and growing
incumbents in the upper mid-band are commercial communication
satellites \cite{huang2020recent}.
Fig.~\ref{fig:satbands} 
shows the standard satellite bands.
We see immediately that the X, Ku, and K
bands -- all widely used by satellites --
fall entirely in the upper mid-band.
In fact, recent FCC reports \cite{fcc2023preliminary,fcc20236Gworkinggroup}
indicate extensive use of the bands
by various commercial satellite services.  
Moreover, with the rapid growth of satellite
Internet services, there is enormous demand
for increased bandwidth, 
particularly in the 7-15\,\si{GHz} bands.
As an example, Fig.~\ref{fig:satbands} shows
the bands granted to the commercial satellite provider Starlink in the recent FCC grant \cite{fcc-starlink-2022}.  
As depicted, the bands in the grant are in both the uplink and downlink, as well
as gateway-satellite and terminal-satellite links.
3GPP has also begun considering 5G NR services
from satellites to mobile devices \cite{3GPP38811,3GPP38821,lin20215g}, 
including communication in the S, K, and Ka bands,
which fall partly within the upper mid-band.
More generally, 
satellite services are growing rapidly
\cite{grandview2023satellite} and if
terrestrial cellular services are to be deployed
in significant fractions of the upper mid-band, co-existence with satellite services will be crucial.

\usetikzlibrary {patterns,patterns.meta}
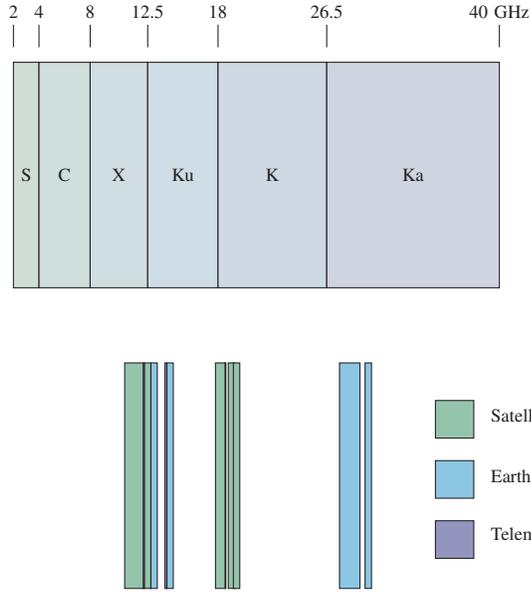
\begin{figure}
    \scriptsize
    \centering 
    \begin{tikzpicture}     

        \foreach \fa/\fb/\col/\text in {2/4/1/S,4/8/2/C,8/12.5/3/X,12.5/18/4/Ku, 18/26.5/5/K, 26.5/40/6/Ka} { 
            \pgfmathsetmacro{\xa}{0.17*\fa};
            \pgfmathsetmacro{\xb}{0.17*\fb};
            \pgfmathsetmacro{\xmid}{(\xa+\xb)/2};
            \pgfmathsetmacro{\huep}{0.3 + 0.4*\col/8};
            \definecolor{cellcol}[rgb]{hsb}{\huep,0.1,0.9}
            
            \filldraw[draw=black,fill=cellcol] (\xa,4) rectangle (\xb,7);
            \draw (\xa,7.2) -- (\xa,7.5) node[above] {\fa};    
            \node[xshift=0cm] at (\xmid,5.5)  {\text};        
        }

        \pgfmathsetmacro{\xb}{0.17*40};
        \draw (\xb,7.2) -- (\xb,7.5) node[above] {40 GHz};    

        \foreach \fa/\fb/\col in { 
            10.7/12.75/1, 17.8/18.6/1, 18.8/19.3/1, 19.7/19.2/1,   
            12.75/13.25/4, 14/14.5/4, 27.5/29.1/4, 29.5/30.0/4,     
            12.15/12.25/7, 18.55/18.60/7, 13.85/14/7     
                        } { 
                \pgfmathsetmacro{\xa}{0.17*\fa};
                \pgfmathsetmacro{\xb}{0.17*\fb};
                \pgfmathsetmacro{\redp}{0};
                \pgfmathsetmacro{\huep}{0.3 + 0.4*\col/8};
                \definecolor{cellcol}[rgb]{hsb}{\huep,0.4,0.9}
                
                \filldraw[draw=black,fill=cellcol] (\xa,0) rectangle (\xb,3);            
            }
        
        \pgfmathsetmacro{\xa}{0.17*35};
        \pgfmathsetmacro{\xb}{0.17*38};
        \pgfmathsetmacro{\xc}{0.17*47};    
        
        \foreach \text/\col/\vert in { Satellite-to-earth/1/2, Earth-to-satellite/4/1.2, Telemetry/7/0.4 }  {
          \pgfmathsetmacro{\huep}{0.3 + 0.4*\col/8};
          \pgfmathsetmacro{\vertb}{\vert+0.5};
          \definecolor{cellcol}[rgb]{hsb}{\huep,0.4,0.9}            
          \filldraw[draw=black,fill=cellcol] (\xa,\vert) rectangle (\xb,\vertb); 
          \node[xshift=0.2cm, yshift=0.3cm] at (\xc,\vert)  {\parbox{3cm}{\text}};
        }
  \end{tikzpicture}
  \caption{Top:  Satellite bands.  Bottom: Recent requests for the Starlink frequency band to
the FCC \cite{fcc-starlink-2022}. 
  Additional requests in the E-Band are not shown.  }
  \label{fig:satbands}
\end{figure}
\begin{figure}
\includegraphics[width=0.9\columnwidth]{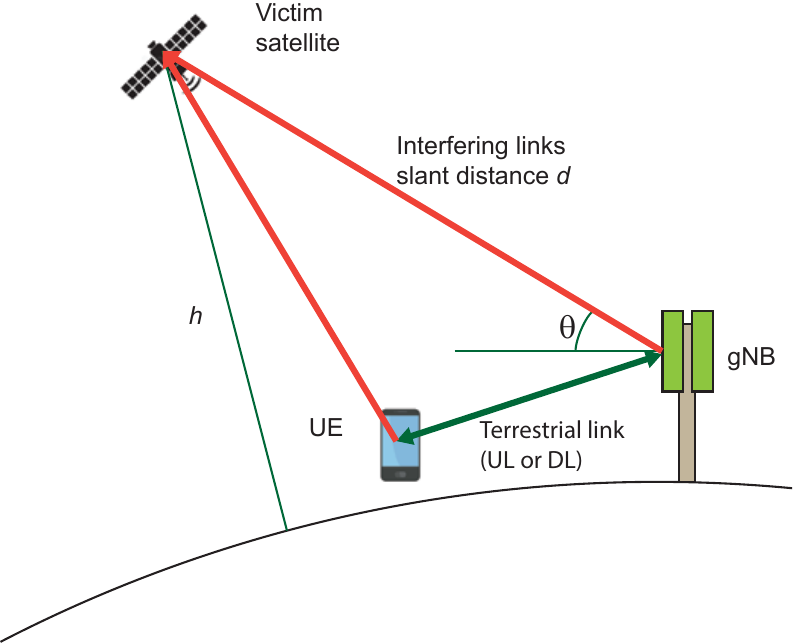}
\caption{Terrestrial to satellite
interference where terrestrial transmissions between a 
terrestrial gNB and UE 
interfere with the satellite uplink (UL).}
\label{fig:sat_int}
\end{figure}

\begin{figure}
\centering
\includegraphics[width=0.95\columnwidth]{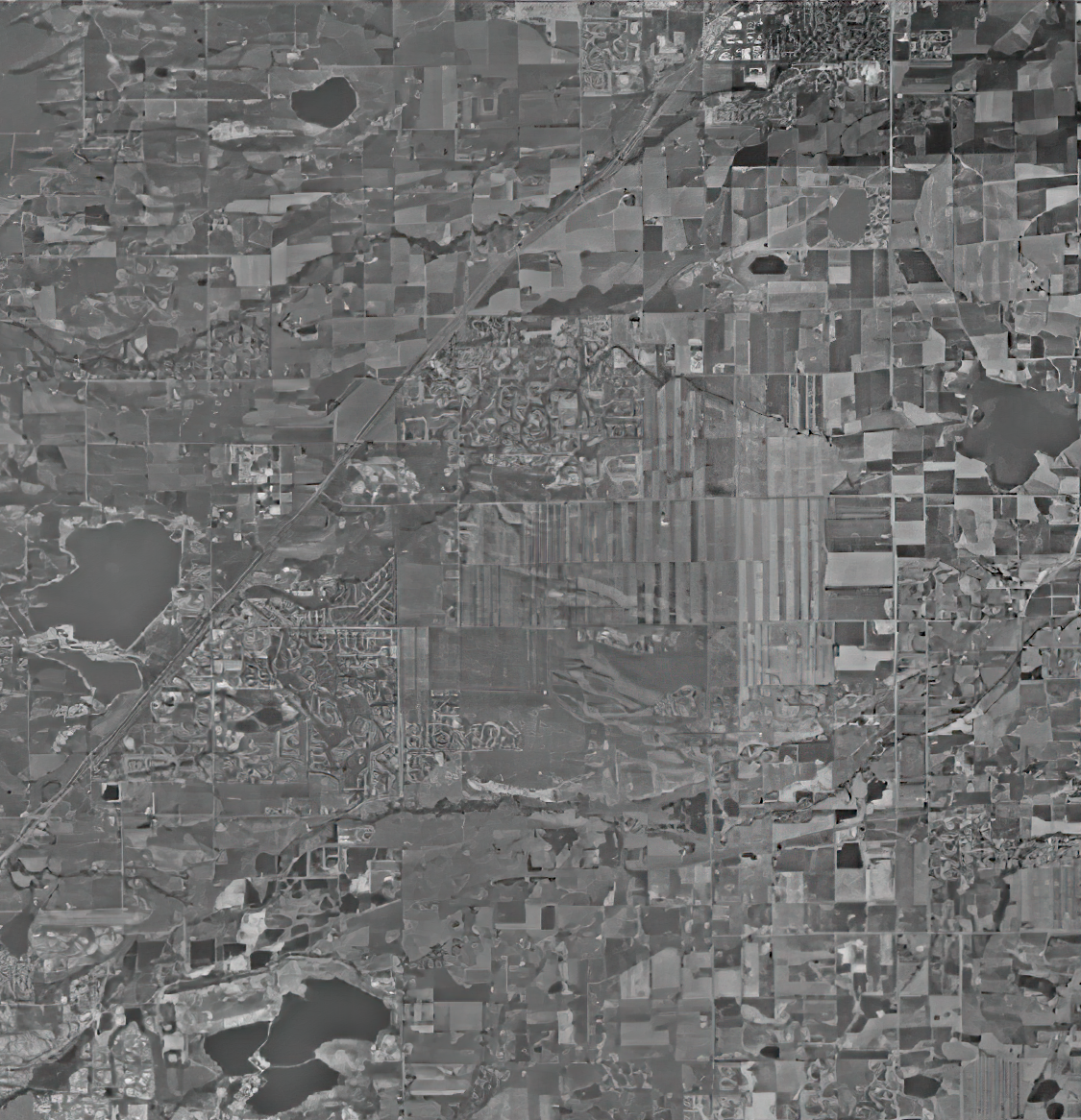}
\caption{\textcolor{black}{Colorado plains captured from ray-tracing simulator for satellite simulation in rural area.}}
\label{fig:rural_area}
\end{figure}

\subsection{Estimating Satellite Interference}
\label{sec:sat_int}

Several recent works have analyzed
interference between satellite and terrestrial networks.  
For example, \cite{sharma2017performance} considers
the selection of the network between satellite and terrestrial services. The works \cite{hattab2018interference,lagunas20205g}
analyze interference of terrestrial networks onto the satellite downlink,
i.e., satellite ground BSs,
in the C-band.  Similarly, \cite{zhang2019spatial}
considers interference on the geostationary (GEO) and medium Earth orbit (MEO) satellite downlink in the Ka band.
The work
\cite{yastrebova2020theoretical} uses stochastic geometry to estimate interference in the satellite uplink, also in the Ka band.

In this work, we estimate the potential interference
between terrestrial cellular networks
and satellite services in the upper mid-band. As illustrated in Fig.~\ref{fig:sat_int}, there exist possible interference channels to the low Earth orbit (LEO) satellite uplink by terrestrial networks.
We will examine both the interference from the terrestrial
downlink (DL, when the gNB is transmitting) and the terrestrial
uplink (UL, when the UE is transmitting). 
When terrestrial UE or gNB transmits data,
some portion of that signal energy is received as interference at a victim
satellite. To obtain multipath channels for the DL and UL cases, we conducted an extensive ray tracing simulation. Based on the ray-tracing data, we build and run a simple system-level simulation to estimate the interference distribution.

\begin{table}[t!]
\footnotesize
  \begin{center}
    \caption{Terrestrial-satellite interference simulation parameters.}
    \label{tab:int_sim}
\begin{tabular}{|>{\raggedright}p{4.2cm}|>{\raggedright}p{3cm}|}

\hline
\cellcolor{purple!15} \textbf{Parameters} & \cellcolor{purple!15} \textbf{Values} 
\tabularnewline \hline 

Carrier frequency $f_c$ & $\{6,18\}$\,\si{GHz} 
\tabularnewline \hline 

{gNB transmit power } $P_{\rm tx}$& \SI{33}{dBm}
\tabularnewline \hline

{UE transmit power } $P_{\rm tx}$ & \SI{23}{dBm} 
\tabularnewline \hline

Satellite transmission bandwidth, $B$ & \SI{30}{MHz} 
\tabularnewline \hline

gNB TX array elevation orientation & $12^\circ$ downtilt
\tabularnewline \hline



Satellite altitude $h$ & \SI{600}{km} 
\tabularnewline \hline

Satellite elevation angle $\theta$ & Uniform in $[10^\circ,90^\circ]$.
\tabularnewline \hline

{Satellite RX $G/T$} & \SI{13}{dB/K} \cite{3GPP38821} 
\tabularnewline \hline

UE and gNB antenna pattern  & 3GPP TR 37.840 \cite{3GPP37840} with $65^\circ$ horizontal and elevation half-power beam width, \SI{30}{dB} front-to-back gain.
\tabularnewline \hline

gNB antenna array  & $8\times 8$ URA 
\tabularnewline \hline


UE antenna array  & $1\times 2$ ULA 
\tabularnewline \hline


Additional propagation losses $L_a$ (gaseous, cloud, fog, 
rain and scintillation)
& \SI{0}{dB}
\tabularnewline \hline

    \end{tabular}
  \end{center}
\end{table}

The system-level simulation parameters are listed in Table~\ref{tab:int_sim}.
Satellite altitude, $h$, and
antenna gain-to-noise-temperature for satellites, $G/T$,
are taken from \cite[Section 6.1.1.1]{3GPP38821} for the Set-1 LEO-600 case in the Ka band.
We presume that the peak gain is achievable in the entire simulation area,
corresponding to the case where we are looking at a satellite beam focused in the simulation region.
Interference in a victim satellite is calculated under the assumption that there is no additional attenuation $L_a$, except for pathloss provided by the ray-tracing simulation, to make the analysis conservative
(including this attenuation will reduce interference).
The antenna pattern for the gNB and UE are taken
\cite{3GPP37840} with half-power
beamwidths of 65$^\circ$ in both azimuth
and elevation, maximum element gain of \SI{8}{dBi},
and front-to-back gain of \SI{30}{dB}.

We selected a typical rural area shown in Fig. \ref{fig:rural_area} and examined a rural setting, as such areas are essential for satellite coverage.
A satellite is placed at an altitude $h=$~\SI{600}{km}
located at a random azimuth angle and an elevation angle $\theta$ uniformly distributed
in $[10^\circ,90^\circ]$. Note that the line-of-sight (LOS) distance to the satellite is the so-called \emph{slant distance} given as
\begin{equation} \label{eq:slant_dist}
    d(\theta) = \sqrt{R_E^2\sin^2(\theta) + h^2 + 2hR_E}
    -R_E\sin(\theta)
\end{equation}
where $R_E$ is the earth radius.  
A gNB and UE are randomly selected in the simulation area with the constraint
that the UE is within \SI{1}{km} of the BS.

 We use ray-tracing data to estimate the multi-input multiple-output (MIMO) channel matrix, $\bsym{H}_{\rm ter}$, between the gNB and UE
and the channel vector, $\bsym{h}_{\rm sat}$, from the gNB to the satellite for the DL case. In the UL case, $\bsym{h}_{\rm sat}$ is the channel vector from the UE to the satellite. In both cases, as further discussed in the next section, $\bsym{h}_{\rm sat}$ can be approximately tracked using ephemeris data.  
For simplicity, we treat the satellite as a single
antenna receiver, since the beamforming gain of the satellite is already incorporated into the $G/T$
value. 
Furthermore, the channel matrices $\bsym{H}_{\rm ter}$ and $\bsym{h}_{\rm sat}$, include the gain of the antenna element for gNB or UE and the multipath components.

In our first simulation, we take into consideration the case where the terrestrial TX and RX select TX and RX beamforming vectors
$\bsym{w}_t$ and $\bsym{w}_r$ to maximize the beamforming gain on the terrestrial link,
\begin{equation} \label{eq:bfopt}
    \wh{\bsym{w}}_t, \wh{\bsym{w}}_r = \arg \max_{\bsym{w}_t, \bsym{w}_r} 
        |\bsym{w}_r^\Herm\bsym{H}_{\rm ter} \bsym{w}_t|^2
\end{equation}
where the optimization is solved over unit vectors.  \textcolor{black}{Assuming TX and RX perform channel estimation using pilot signals}, the solution to \eqref{eq:bfopt} is given by
the maximum singular vectors of $\bsym{H}_{\rm ter}$ \cite{heath2018foundations}.

Importantly, the selection of the beamforming vectors from (\ref{eq:bfopt}) do not take into account the interference to the satellite uplink -- it only maximizes the SNR on the terrestrial link.
After taking the TX beamforming vector $\wh{\bsym{w}}_t$, 
the resulting channel from terrestrial TX to satellite RX will be $\wh{\bsym{w}}_t^\Herm \bsym{h}_{\rm sat}$.
Hence, \textcolor{black}{referencing the definition of carrier-to-noise ratio \cite{3GPP38811}}, the interference-to-noise ratio (INR) in dB at the satellite will be
\begin{align}
    \MoveEqLeft \INR = P_{\rm tx} + 10 \log_{10} |\wh{\bsym{w}}_t^\Herm \bsym{h}_{\rm sat}|^2 
    \nonumber \\
    &+ \frac{G}{T} - L_a - 10\log_{10}(B) - 10 \log_{10}(k)
\end{align}
where $P_{\rm tx}$ is the total TX power
of the UE or gNB, 
$G/T$ is the satellite RX gain to thermal noise ratio in dB,
$L_a$ are the other propagation losses, 
$B$ is the bandwidth, and $k$
is Boltzmann's constant. 
Note that $\INR$ is a valuable metric for assessing the interference penalty in satellite services. 
 Specifically, given an INR value,
an intended uplink signal to the satellite will experience a degradation in SNR of
\begin{equation}
    \Delta = 10 \log_{10}(1 + 10^{0.1\cdot\INR})
\end{equation}
Typical satellite systems require an $\INR <$\,\SI{-6}{dB} corresponding to an SNR
degradation of  $\Delta \approx$\,\SI{1}{dB}. 


\begin{figure}[t]
  \centering
  \includegraphics[width=0.99\columnwidth]{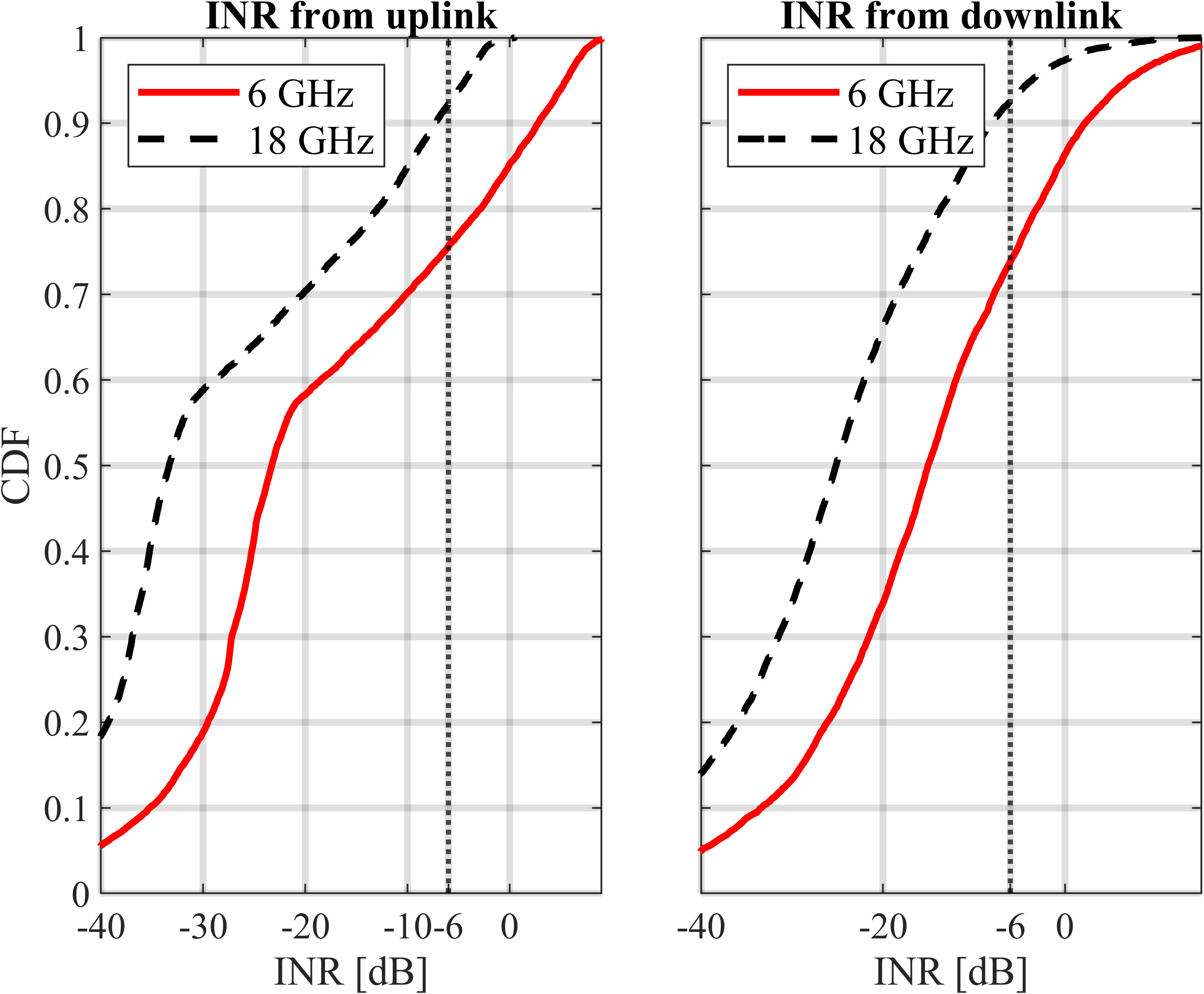}
  \caption{INR distribution from UL and DL terrestrial cellular sources. }
  \label{fig:inr}
\end{figure}

Plotted in Fig.~\ref{fig:inr} are the CDFs of the $\INR$ for both the uplink and downlink at two upper mid-band
frequencies:  $f_c=$ 6 and 18\,\si{GHz}.  There are three important conclusions:
\begin{itemize}
    \item \emph{Possibility of high interference to the satellite uplink:}  We see that 
    the $\INR$ can be high.  For example, at $6$ GHz carrier frequency we see that approximately $27$\% of the
    downlink transmissions result in an $\INR \geq -6$\,\si{dB}, the level at which the satellite SNR
    degrades by more than $\Delta=$\,\SI{1}{dB}.

    \item \emph{Both uplink and downlink terrestrial interference sources can be significant:}  We see that the $\INR$s can be large for transmissions by both the UE and the gNB.  
    Although the UE transmits with a lower power, 
    its antenna arrays can be in an arbitrary orientation. Therefore, with some probability, it can be oriented directly to the satellite. In the left panel of Fig.~\ref{fig:inr}, 
    we see that, even with terrestrial uplink transmissions, the $\INR >$\,\SI{-6}{dB}
    about 25\% of the time when $f_c=$\,\SI{6}{GHz}. In contrast, the gNB is equipped with fixed downtilted antenna arrays. In such a case, the interference to the satellite uplink is caused by sidelobes occurring due to transmit beamforming of gNB.
    \item \emph{Reduced interference at higher frequencies:}  By Friis' law, the $\INR$s are lower
at higher frequencies.  This property is considered to be useful for frequency adaptation to reduce interference to satellite services.
\end{itemize}

\subsection{Reducing Satellite Interference with Nulling}
To mitigate the interference, the terrestrial transmitter can employ interference nulling as
follows: Assume, for the time being,
that the terrestrial transmitter knows the channel vector 
$\bsym{h}_{\rm sat}$ 
to the satellite along with the MIMO channel matrix $\bsym{H}_{\rm ter}$ between the gNB and the UE.
{We will discuss the estimate of the satellite channel  vector in the next sub-section.}
As before, the receiver selects the beamforming vector $\wh{\bsym{w}}_r$ from \eqref{eq:bfopt}. However, 
to eliminate interference,
the transmitter selects a vector $\bsym{w}_t$ via the regularized cost:
\begin{equation} \label{eq:bflam}
    \wh{\bsym{w}}_t^\lambda = \arg \max_{\bsym{w}_t} \left[ 
        |\bsym{w}_r^\Herm\bsym{H}_{\rm ter} \bsym{w}_t|^2 - \lambda |\bsym{w}_t^\Herm \wh{\bsym{h}}_{\rm sat}|^2
        \right]
\end{equation}
where $\lambda > 0$ is a regularization parameter.  The regularization term, 
$|\bsym{w}_t^\Herm \bsym{h}_{\rm sat}|^2$, penalizes the interference in the satellite receiver and
attempts to create a null along 
the satellite channels. 
The solution $\wh{\bsym{w}}_t^\lambda$ to \eqref{eq:bflam} is given by maximum eigenvector of 
\begin{equation}\label{eq:bflam_solt}
\begin{aligned} 
{\bsym{H}}^\Herm_{\rm ter}\bsym{w}_r\bsym{w}_r^\Herm{\bsym{H}}_{\rm ter} - \lambda {\bsym{h}}_{\rm sat}{\bsym{h}^\Herm_{\rm sat}} 
\end{aligned}
\end{equation}

\begin{figure}[t]
  \centering
  \includegraphics[width=0.99\columnwidth]{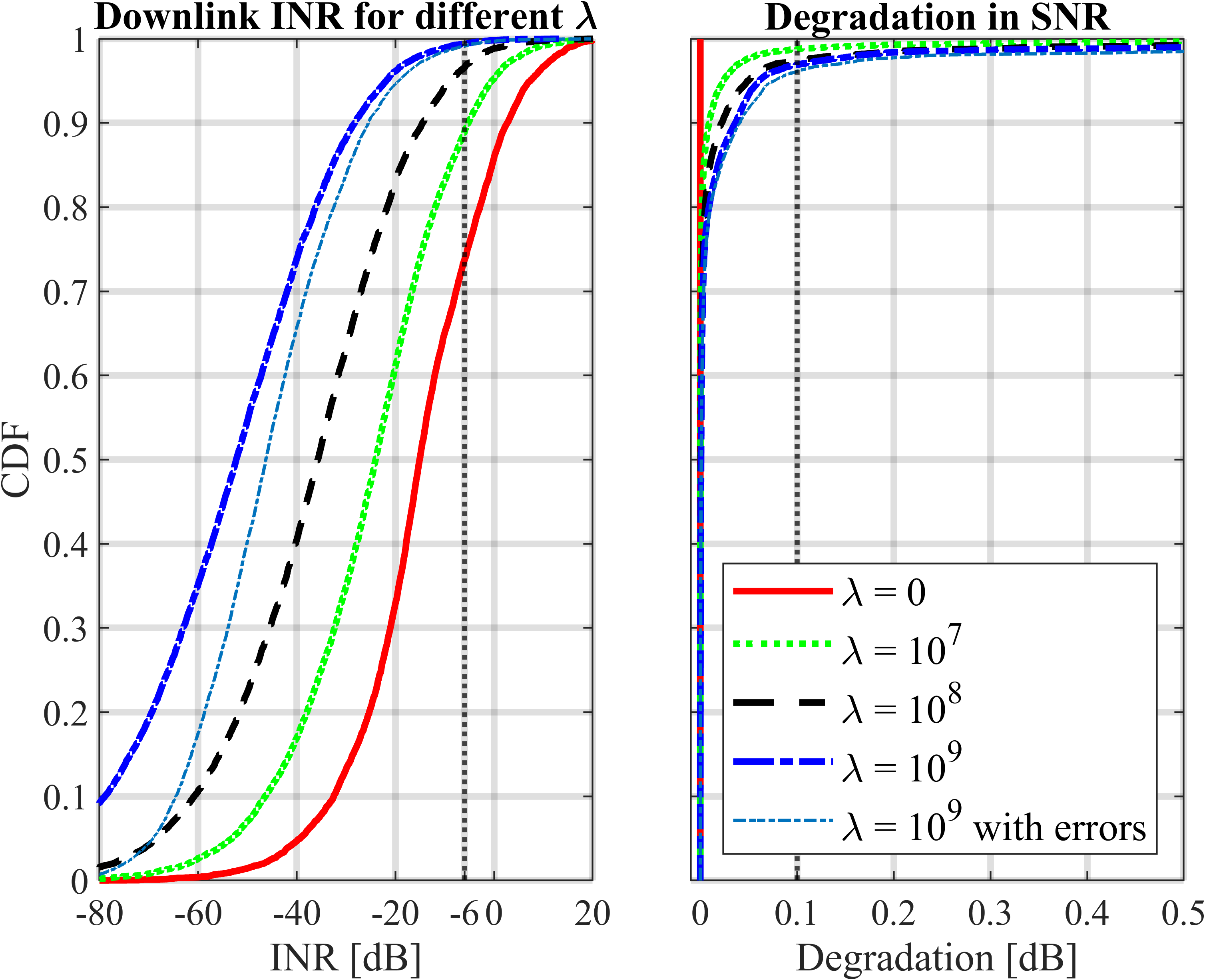}
  \caption{INR distribution from DL terrestrial cellular sources for different $\lambda$ values at a carrier of \SI{6}{GHz}.
  }
  \label{fig:inr_gainloss_dl}
\end{figure}

\begin{figure}[t]
  \centering
  \includegraphics[width=0.99\columnwidth]{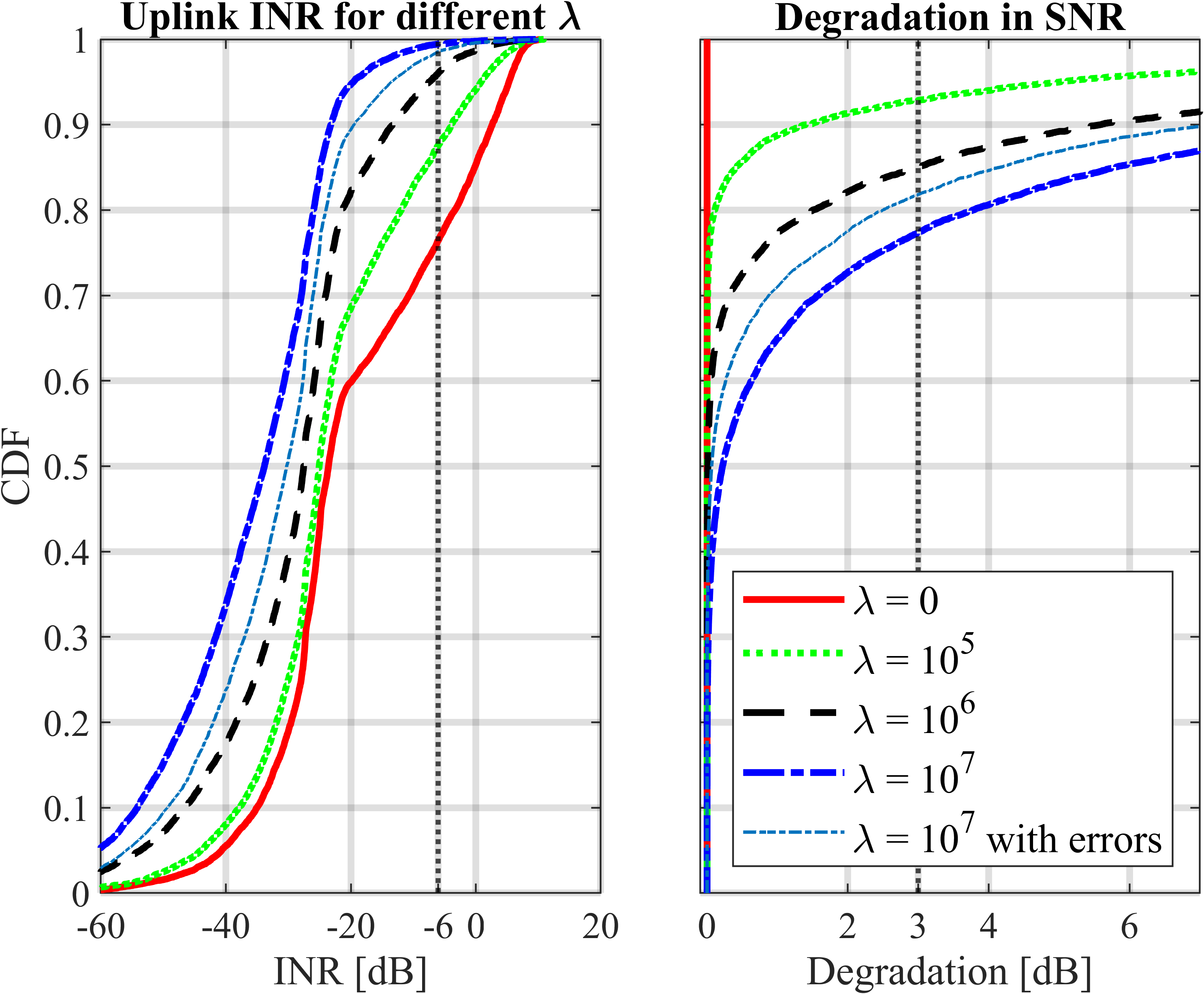}
  \caption{INR distribution from UL terrestrial cellular sources for different $\lambda$ values at a carrier of \SI{6}{GHz}.
  }
  \label{fig:inr_gainloss_ul}
\end{figure}

To assess the effectiveness of the nulling, 
we re-ran the identical simulation as in the previous
subsection where the TX beamforming vector
is computed from \eqref{eq:bflam_solt}.
We have considered the \SI{6}{GHz} carrier frequency
since the interference is higher in the
lower band. 
The left panel of Fig.~\ref{fig:inr_gainloss_dl} 
plots the CDF of $\INR$ caused by terrestrial downlink transmissions to the satellite uplink
at different values of $\lambda$.
{Regarding the curves labeled ``with errors", we will discuss them in the next
sub-section}.
From the left panel of Fig.~\ref{fig:inr_gainloss_dl},
we see that with sufficiently
high $\lambda$, the interference can be well mitigated, assuming ideal tracking. For example,
when $\lambda = 10^{8}$, $\INR \geq $\,\SI{-6}{dB}
less than 3\% of the time.  

Plotted on the right panel of Fig.~\ref{fig:inr_gainloss_dl} is the degradation in the SNR on the terrestrial link defined by
\begin{equation}
    \rho = 10 \log_{10}\left( \frac{|\wh{\bsym{w}}_r \bsym{H}_{\rm ter} \wh{\bsym{w}}_t|^2}{
    |\wh{\bsym{w}}_r \bsym{H}_{\rm ter} \wh{\bsym{w}}_t^\lambda|^2} \right)
\end{equation}
which is the ratio of the beamforming gain with the optimal beamforming vector $\wh{\bsym{w}}_t$ 
from \eqref{eq:bfopt} and the beamforming gain with the regularized beamforming vector $\wh{\bsym{w}}_t^\lambda$
from \eqref{eq:bflam}.  We see that with
a value of $\lambda=10^8$, the
degradation of the terrestrial SNR can be kept
at $\rho < $\SI{0.1}{dB} for 97\% of the time.

Fig.~\ref{fig:inr_gainloss_ul} similarly plots
the CDF of the INR for terrestrial uplink transmission under different values $\lambda$
along with the terrestrial degradation of the SNR $\rho$. 
We see here that the UEs can also
reduce the interference, but the degradation is higher on the terrestrial link. We observe that the degradation of the SNR is less than \SI{3}{dB} for $78\%$ of time when $\lambda = 10^7$.
This arises from the fact that the UE has a lower number of antennas, and hence
directional nulling is more costly.

{
\subsection{Tracking Interference Channels to Satellites} 
A practical issue with the above interference nulling method is that the 
UE and gNB must track the wideband channels to the satellite.  
Specifically, the cellular transmitter
(UE or gNB) must estimate the
channel $\bsym{h}_{\rm sat}$ at all frequencies.
In this sub-section, we assess the possibility of using
satellite \emph{ephemeris data} to track azimuth and zenith angles of satellites.
Ephemeris data are typically publicly available --- see, for example, ~\cite{celesTrack}.
The use of these data has been considered by 3GPP
~\cite{3GPP38821}.
}


{To realistically assess the tracking with the ephemeris data,
we need to take into account several practical 
challenges.  First, knowing the satellite locations, one can only compute the LOS interference channels for victim satellites.  Therefore, we can use the ephemeris data to estimate the LOS component of the interference channel, ignoring the NLOS components.
Since we are considering a rural scenario, the channels are dominated by LOS components, so considering only LOS interference channels should not significantly degrade performance.
We will verify this assumption in the 
simulation below.
}

\begin{table}[t!]
\footnotesize
  \begin{center}
    \caption{Angular errors for tracking experiment.}
    \label{tab:angerror}
\begin{tabular}{|>{\raggedright}p{1.5cm}|>{\raggedright}p{2.7cm}|>{\raggedright}p{1.3cm}|
>{\raggedright}p{1.3cm}|}

\hline
\cellcolor{purple!15} \textbf{Transmitter} &
\cellcolor{purple!15} \textbf{Orientation device} &
\multicolumn{2}{c|}{\cellcolor{purple!15} \textbf{RMS angle error}} 
\tabularnewline \cline{3-4}
\cellcolor{purple!15} &
\cellcolor{purple!15} &
\cellcolor{purple!5} Azimuth &
\cellcolor{purple!5} Elevation 
\tabularnewline \hline

gNB & High-precision compass, e.g.~\cite{ec_compass}
& 0.3$^\circ$ & 0.1$^\circ$
\tabularnewline \hline

UE & 3D magentometer with IMU and filtering, e.g.~\cite{teufl2018validity}
& 1.5$^\circ$ & 1.5$^\circ$
\tabularnewline \hline
\end{tabular}
  \end{center}
\end{table}

{
A second issue is the orientation calibration for ground antenna arrays. 
To use the ephemeris data, the orientation
of the gNB or UE relative to the global frame 
must be known.  For the gNB (base station),
the orientation can be measured, for example,
with a high-precision compass\cite{ec_compass}.  
For the UE, the orientation can be measured with a magnetometer along with an inertial motion unit (IMU) and filtering techniques for tracking.  
These systems will introduce some angular error.  Even if small, performance could be impacted, since spatial nulling tends to create very sharp nulls.
To see this effect, we simulate the
angular errors shown in Table~\ref{tab:angerror} taken
from the published measurements in
\cite{ec_compass,teufl2018validity}.
Specifically, we applied a random angular error given by the values in Table~\ref{tab:angerror} to determine the estimated interference channels $\bsym{h}_{\rm sat}$ towards the victim satellites.
}

{
To better account for the angular errors,
we modified the transmit beamforming procedure as the following
\begin{equation} \label{eq:bflam_robust}
    \wh{\bsym{w}}_t^\lambda = \arg \max_{\bsym{w}_t} \left[ 
        |\bsym{w}_r^\Herm\bsym{H}_{\rm ter} \bsym{w}_t|^2 - \lambda \mathbb{E}|\bsym{w}_t^\Herm  \bsym{h}_{\rm sat}|^2
        \right]
\end{equation}
where $\bsym{h}_{\rm sat}$ includes only LOS component of interference channel to a satellite. 
The optimization is identical to \eqref{eq:bflam}
except for the regularization term
$\mathbb{E}|\bsym{w}_t^\Herm  \bsym{h}_{\rm sat}|^2$.  
The expectation is taken over
the distribution of channel vectors
$\bsym{h}_{\rm sat}$ given the measured angular errors.
This expectation encourages the creation of nulls,
not just for the estimated angles of the satellites
but rather for small regions around those angles to account
for angular errors. The solution to \eqref{eq:bflam_robust} is described in a way similar to \eqref{eq:bflam_solt}, but the expectation for $\bsym{h}_{\rm sat}\bsym{h}_{\rm sat}^\Herm$ should be taken over the distribution of the channel vectors $\bsym{h}_{\rm sat}$.
}

{
The final potential issues are variation of the tracked satellite location over the transmission time interval (TTI) and time synchronization.  
These time-related errors are, however, negligible. For example, since the TTI in 5G is generally less than \SI{1}{ms}, and the velocity of a LEO satellite at an altitude of $h=$\SI{600}{km} is approximately $v=\,$\SI{7.56}{km/s},
the maximum angular variation is at most $\Delta \theta = \tan^{-1}\left( {vT}/{h}\right) \approx 7.2\times 10^{-4}$ degrees. This error is considerably less than the angular errors from the orientation calibration and can thus be neglected.  
The variation in tracked information due to synchronization errors in gNBs is also negligible, as network synchronization protocols such as the Precision Time Protocol (PTP) in the open radio access network (O-RAN) require nanosecond-scale timing synchronization errors \cite{3GPP103859}.
}


{Under these assumptions, we evaluate the modified interference nulling scheme given in \eqref{eq:bflam_robust} by running the same simulation as in the previous section.  Fig.~\ref{fig:inr_gainloss_dl}, shows the INR and gain loss CDFs with angular errors for the DL on the curve labeled
``$\lambda = 10^9$ with errors".  We see that the performance is almost identical to the curve without errors.  
Similarly, as shown in the curve labeled
``$\lambda = 10^7$ with errors" of Fig.~\ref{fig:inr_gainloss_ul}, we see that there is no significant change of INR due to angular errors for the UL case.
Note that while tracking errors for UEs
are assumed to be larger, as shown in Table~\ref{tab:angerror},
the effect of angular error is not as strong, as UEs have fewer antennas and produce
less narrow nulls. 
Furthermore, since the nulls are created only on the LOS component, in the right panel of Fig.~\ref{fig:inr_gainloss_dl}, we see the lower loss in SNR compared to the case without errors. 
For both gNB and UE, these results suggest that the modified interference nulling technique can
work with reasonable tracking errors, even for high-speed LEO satellites.
}

\subsection{Radio Astronomy} 
Radio astronomy studies the electromagnetic emissions from distant astronomical sources and high energy events \cite{condon2016essential,burke2019introduction}. These emissions are wideband in nature, and reach the surface of the Earth strongly attenuated by their long-distance propagation throughout the interstellar or intergalactic media, but also due to atmospheric absorption.

Astronomical emissions can be characterized by their spectral shapes. They feature persistent continuum emissions depending on their nature \cite{national2010spectrum,national2007handbook}: blackbody radiations for the cosmic background, free-free emissions, e.g. for star-forming regions, or synchrotron emissions e.g., for neutron stars.

Gaseous or ionized sources also embed narrow features known as emission or absorption spectral lines \cite{rohlfs2013tools,barrett1958spectral}. These lines are characteristics of the chemical elements present in the astronomical source, and are used to trace their composition, structure, and density. They are the only probes available to the interstellar medium and to external galaxies. They also reveal complementary information, such as gas temperatures, ionization, and fluid dynamics. More importantly, their shift in frequency from a given rest frequency, known as \textit{redshift}, provides information on the age of the source and its distance to an observer. The International Astronomical Union (IAU) defined a list of the most important spectral lines, in which more than 25\% fall in and below the mid-band ($<$\,\SI{24}{GHz}). Notably, the complex prebiotic molecules, essential to the understanding of life processes in the Universe, have spectral signatures concentrated between 10 and 15\,\si{GHz} \cite{thaddeus2006prebiotic,corby2016astrochemistry}.

The spectral flux density of an astronomical sources is expressed in jansky (\si{Jy}), which is defined as \SI{1}{Jy} = $10^{-26}$ Wm$^{-2}$Hz$^{-1}$.   To appreciate
how small this flux density is, recall that
an isotropic antenna at $f_c=$\,\SI{6}{GHz}
would have an aperture of $A = \lambda^2/(4\pi) \approx 2(10)^{-4}$\,\si{m^2}.
A signal of \SI{1}{Jy} with this antenna 
would thus be received at \SI{-267}{dBm},
more than \SI{90}{dB} below the noise floor.
Observing milli- to micro-Jy sources are not uncommon with modern radio telescopes.

Detecting these weak emissions requires high sensitivity, which is achieved with radio telescopes using large collecting areas from wide reflectors or combined across multiple dish antennas, receivers with low system temperatures and wide bandwidths, and long integration times spanning seconds to hours of data integration.

Radio telescopes are also sensitive to Radio Frequency Interference (RFI), which can impact astronomical observations at various levels. Weak sources of RFI are detected after data integration in either or both time and frequency domains, and the associated corrupted time and frequency resource blocks are then discarded before further astronomical information extraction processes \cite{baan2011rfi,ford2014rfi}. The loss of data associated with this procedure not only reduces the sensitivity of an astronomical observation, which can possibly be recovered by longer observations, but may also prevent the observation of transient sources, such as Fast Radio Bursts \cite{petroff2019fast} or counterparts of sources of gravitational waves, such as super massive black holes mergers \cite{hotokezaka2016radio}, which are individual and non-repeatable events. Similarly, redshifted spectral lines can fall outside protected frequency bands where the astronomical information may be fully lost. This is for instance the case with the redshifted galactic Hydrogen line with rest frequency at 1400 MHz falling into the lower Global Navigation Satellite System (GNSS) bands (1145-1310 MHz), and preventing the observation of the edges of the local Universe \cite{gilloire2009rfi}. Stronger sources of RFI can drive the electronics components of a telescope receiver into a nonlinear regime, leading to the complete loss of data loss.

{The impact of RFI is minimized by locating radio observatories in remote areas with low population and terrestrial transmitter densities, exploiting the propagation losses of these transmissions due to their large distance from a radio observatory \cite{umar2014importance}. Further protection can be sought through coordination with active services to prevent the deployment of future radio frequency infrastructure, as is the case in the National Radio Quiet Zone (NRQZ) \cite{sizemore2002national,series2021characteristics}. Emerging work involving artificial intelligence (AI) and machine learning (ML) based nonlinear signal processing may help partially offset the effects of RFI from LEO constellations  on radio astronomy \cite{comcas23madanayake}.}

\section{Wideband Antennas for the Upper Mid-Band}
\label{sec:antenna}

\subsection{Challenges}
The antennas and RF circuits of
upper mid-band transceivers need to support wide bandwidths, high degree of tunability, and large numbers of antenna 
elements for directionality.  Existing technologies have a number of bottlenecks  in meeting these requirements including: a) antenna size-bandwidth-gain tradeoffs, 2) high frequency losses, 3) degraded SNR while scanning, 4) poor interference tolerance, and 5) inefficient spectrum management. To address the above shortfalls, here we combine a mix of wideband elements that are tunable and reconfigurable to mitigate interference and improve SNR across dynamically large spectrum swaths in the upper mid-bands.

 Even though an antenna element can be designed to operate across the full range of interest, there are trade-offs that limit performance/ sensitivity and directional gain as a strong function of frequency if a single element is used in an array configuration for sensing applications. We want to avoid such compromises and trade-offs while covering the full band of interest, which requires several innovations in the antenna and microwave circuit areas for a cost-effective and scalable high-performance solution. Another key challenge in developing wideband systems for cellular applications is the antenna form factor.  While there has been extensive work in wideband antennas (see, for example, \cite{lim2010ultra}), most designs require physically thick profiles that are not suitable for portable devices.
 


\subsection{Compact Wideband Aperture in Aperture Antenna Arrays}

\begin{figure}
\centering
\centering
\includegraphics[width=\columnwidth]{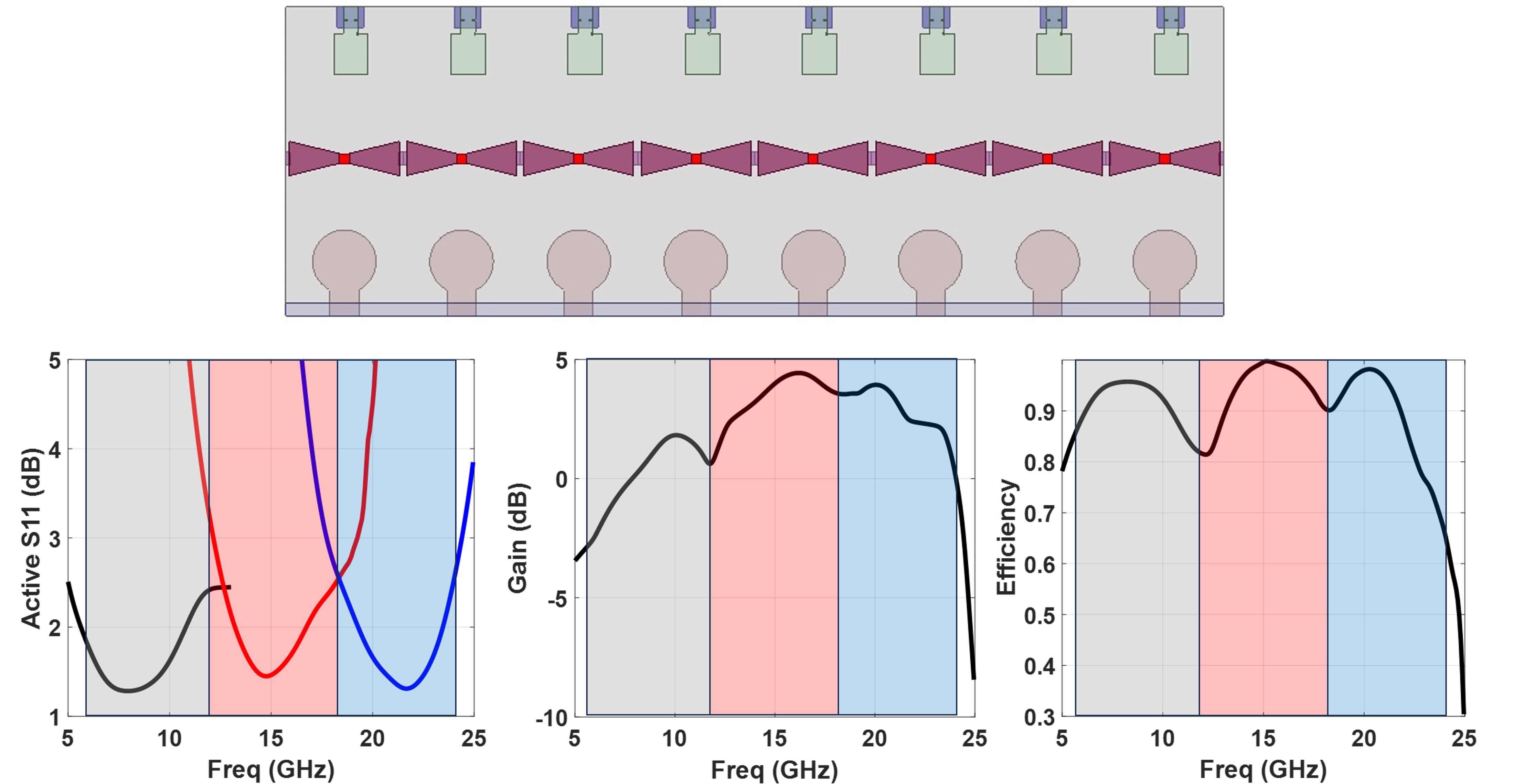}
\vspace{-.2in}
   \caption{{\textcolor{black}{(Top) HFSS illustration of tri-band array, (Bottom) Simulated plots showing active VSWR, gain and efficiency of the array when all ports are excited.}} }\label{AIA}
\vspace{-.1in}
\end{figure}

  To overcome the aforementioned challenge, a low-profile aperture-in-aperture (AiA) realization is investigated, considering 3 classes of antennas: a) ultra-wideband (UWB) tightly coupled arrays (TCDAs) \cite{TCDA1,TCDA2} \textcolor{black}{\cite{Doane_TCDA,ZandermmW,tzanidis1,Jingni_TCDA}}, b) low profile planar circular monopole \cite{Mono1,Mono2} \textcolor{black}{\cite{Mono3,Mono4,Mono5}}, and c) UWB patches \cite{Patch1, Patch2}. \textcolor{black}{\cite{Patch3,Patch4}}. \textcolor{black}{Specifically, the unit cell of this array (see Fig.~\ref{AIA} was designed based on existing work in literature, which was later optimized to meet the requirements of the FR3 bands. Notably, the UWB performance offered by the TCDA is based on Wheeler's current sheet principle. In addition, the ground plane inductance is canceled using the capacitive overlaps of the dipoles, thereby eliminating unwanted resonances at certain frequency spots \cite{Doane_TCDA,ZandermmW}. With the monopole, the ground plane, which serves as the counterpoise for the monopole, is designed to enhance the antenna's performance {\cite{Mono3,Mono4,Mono5}}. The size and shape of the ground plane play a role in determining the bandwidth. Similarly, by expanding the above approach, the bandwidth of the patch antenna was also increased. That is, the edges of the partial ground plane of the patch antenna were modified to allow for radiation, resulting in improving the impedance bandwidth of the patch antenna. Of course, this list is not exhaustive, and there are other types of wideband antennas that can also be considered. For instance, the holographic antennas can give the required bandwidth, as the holographic technique allows the use of sub-wavelength unit cells, and these reflectarrays and transmitarrays, such as \cite{Holo1,Holo2,Holo3}, can be used in cellular network towers.} 
  
  Notably, this AiA brings forward several unique features with game-changing impact in antenna array design and performance features. Specifically, they are: 1) scalable across all frequencies and geometries; 2) highly compact in terms of element size and thickness; 3) wide and continuous bandwidth of more than; 4) low-cost and easy to deploy in a highly conformal manner to mounted and deployed easily on any platform. Indeed, realizing low-cost fabrication and beamforming across wide bandwidths is one of the foundational challenges in our footsteps. The AiAs overcome this challenge quite effectively and with all the required characteristics rather naturally. 

\begin{figure}
\centering
\centering
\includegraphics[width=\columnwidth]{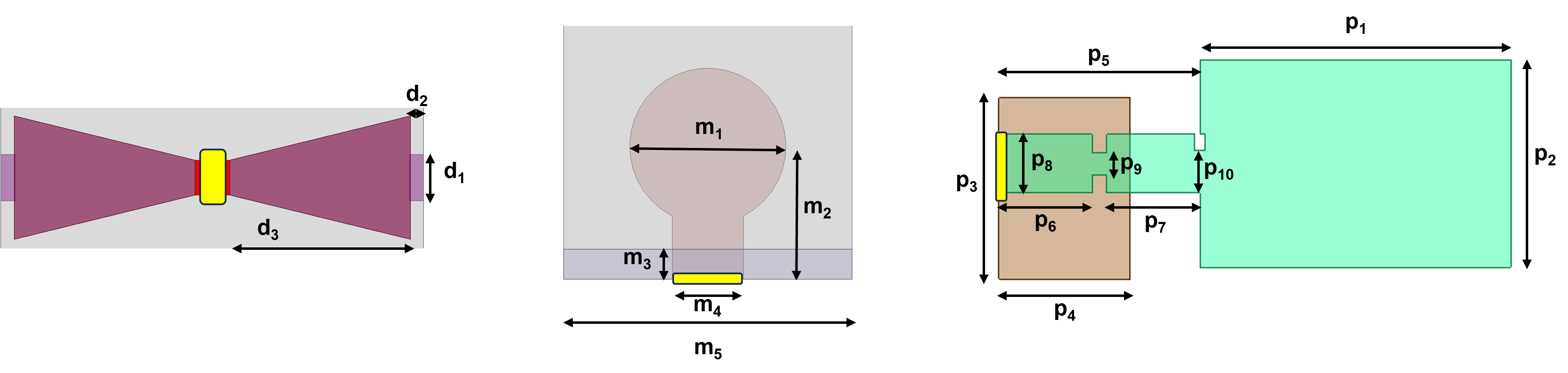}
\vspace{-.2in}
   \caption{{Dimensions of tri-band antennas. Excitation ports are depicted in yellow. } }\label{Dim}
\vspace{-.1in}
\end{figure}

Unlike existing UWB arrays which use a single aperture for the entire band, here we present three separate antenna designs for each band. Doing so, provides the desired beam scanning across all the bands which is not the case with the former. Due to finite array size, using a single aperture significantly limits the beamscanning performance of the array due to the finite electrical array size at the lower bands. For instance, a typical $8 \times 8$ UWB array operating across a 4:1 impedance bandwidth, will have an antenna element spacing of $\lambda_{\rm high}/2$. This corresponds to an inter-element spacing of $\lambda_{\rm low}/8$ at the lowest band, implying that the effective number of elements at the lower band (6 GHz) is only 2. This significantly impacts the array scanning performance. 

 \begin{figure}
\centering
\centering
\includegraphics[width=\columnwidth]{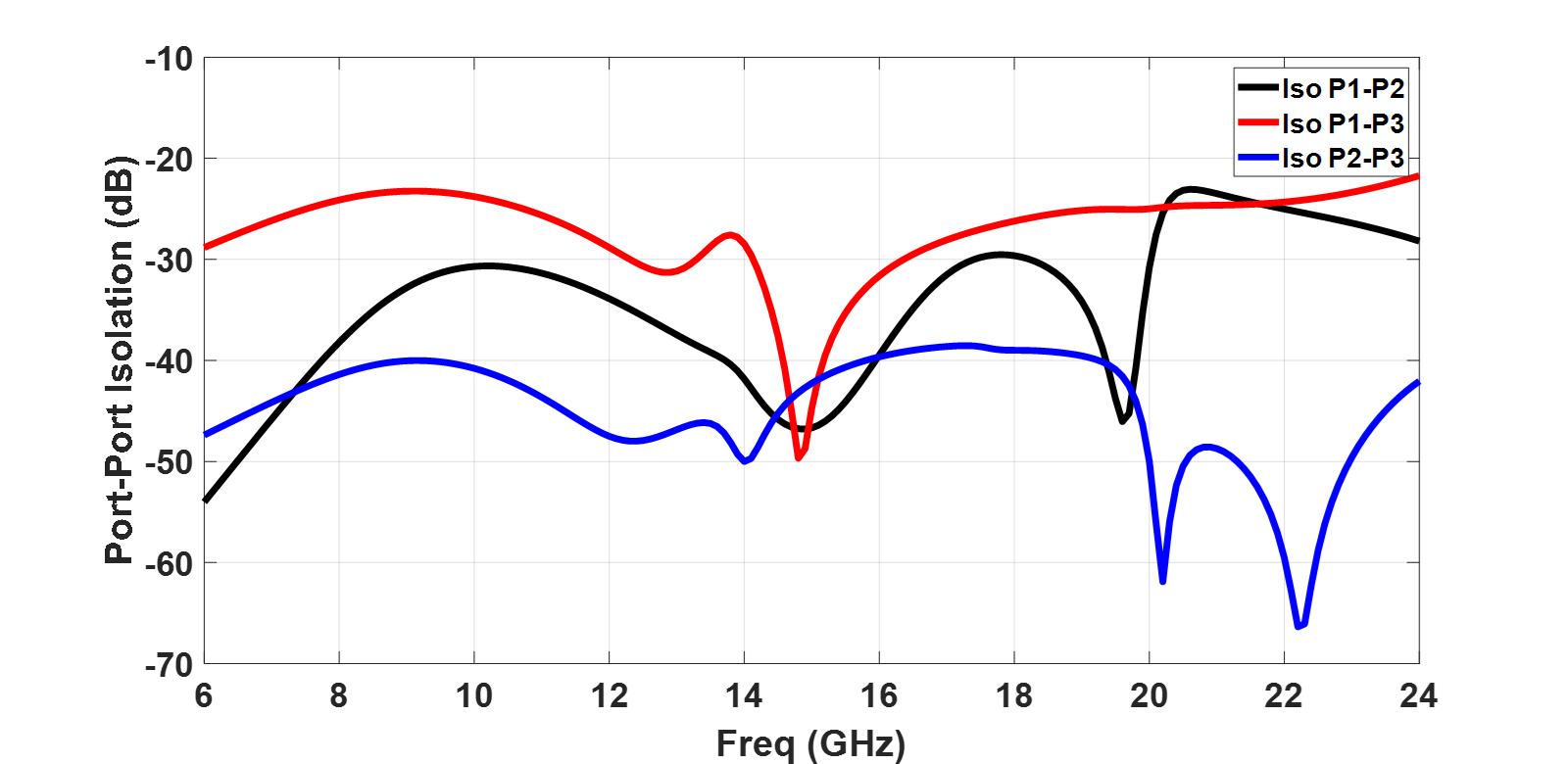}
\vspace{-.2in}
   \caption{{Port-port isolation exhibited by the co-located tri-band antennas. P1 corresponds to 6 - 12 GHz dipoles, P2 monopole spans 12 - 18 GHz, while P3 is represented by the patches operating across 18 - 24 GHz.} }\label{Iso}
\vspace{-.1in}
\end{figure}

 \begin{figure}
\centering
\centering
\includegraphics[width=\columnwidth]{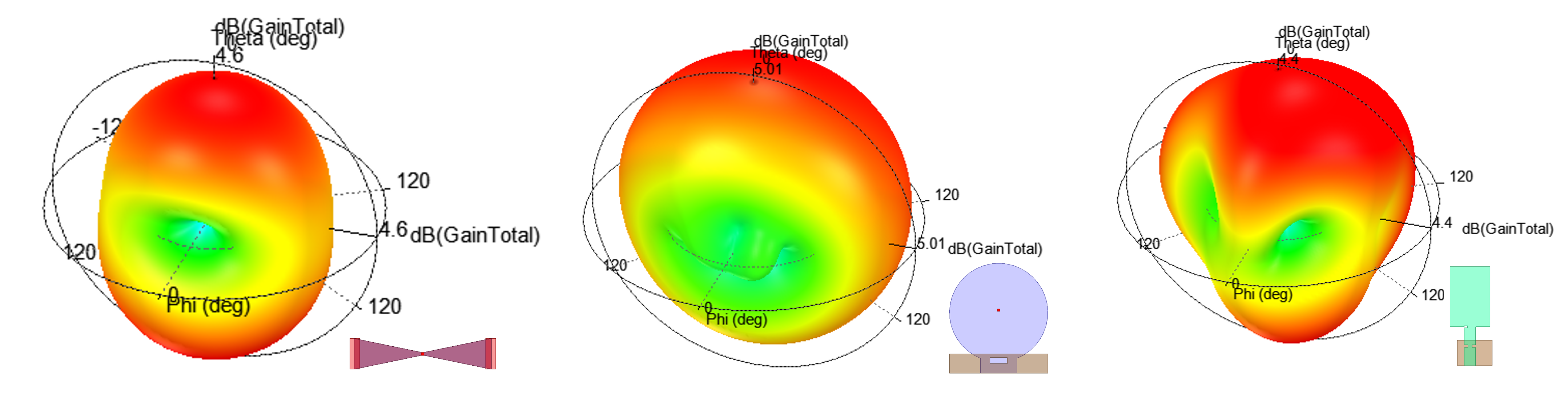}
\vspace{-.2in}
   \caption{{Unit cell radiation pattern at the center frequency of all 3 bands.} }\label{Rad}
\vspace{-.1in}
\end{figure}

\begin{table}[t]
\footnotesize
  \begin{center}
    \caption{Tri-band Antenna unit cell dimensions (in mm).}
    \label{T3}

\begin{tabular} { | p {1.5 cm} | p {1.5 cm} | p {1.5 cm} | p {1.5 cm} | }

\hline 

\cellcolor{purple!15} \textbf{Dimension} & \cellcolor{purple!15} \textbf{Value [mm]} &
\cellcolor{purple!15} \textbf{Dimension} & \cellcolor{purple!15} \textbf{Value [mm]}
\tabularnewline \hline

$d_1$ & 1.125 &  $p_2$ & 3
\tabularnewline \hline

$d_2$ & 0.35 & $p_3$ & 2.35
\tabularnewline \hline

$d_3$ & 4.4 & $p_4$ & 1.79
\tabularnewline \hline

$m_1$ & 5.6 & $p_5$ & 2.47
\tabularnewline \hline

$m_2$ & 4.8 & $p_6$ & 1.1 
\tabularnewline \hline

$m_3$ & 1.07 & $p_7$ & 1
\tabularnewline \hline

$m_4$ & 2.4 & $p_8$ & 1.3 
\tabularnewline \hline

$m_5$ & 10.25 & $p_9$ & 0.715
\tabularnewline \hline

$p_1$ & 3.6 & $p_{10}$ & 1.18
\tabularnewline \hline








    \end{tabular}
  \end{center}
\end{table}

To address the above shortfall, a single aperture-in-aperture (AIA) antenna array comprising of 3 separate antenna arrays is designed. The lower band from $6 - 12$\,\si{GHz} will comprise of closely spaced dipole arrays employing a co-axial feed, the mid-band from  $12-18$\,\si{GHz} will employ circular monopole, while the high-band from $18-24$\,\si{GHz} will be realized using patches. The top panel of Fig.~\ref{AIA} depicts a simple illustration of the AiA geometry. The unit-cell dimensions are provided in Table.~\ref{T3}. Notably, Ro4003 with permittivity ($\epsilon = 3.38$), loss tangent ($tan \delta = 0.0027$) was employed as a substrate with a thickness of \SI{0.813}{mm}. \textcolor{black}{ The constant spacing between elements has been set to be 9 mm. Thus, the entire 8-element array measures 74 mm $\times$ 25 mm.}

As depicted, a single aperture houses antennas operating across the entire upper mid-band band. \textcolor{black}{The design optimization and simulation were performed using commercially available Ansys HFSS. The unit cell comprises of one antenna of each type with a semi-infinite boundary condition set-up. Hence, we used only 3 ports for our simulation depicted in yellow in Fig.~\ref{Dim}.  Active unit-element voltage standing wave ratio (VSWR), gain, and efficiency are provided in the bottom panel Fig.~\ref{AIA}. As shown, all three antennas have a very good return loss or VWSR $< 3$. Further, the unit cell gain ranges from -2 to 4 dB across the operational FR3 bands. Finally, the average efficiency of this aperture is 82\% with the upper bands experiencing a minimum efficiency of 67\% at the upper ends. } Fig.~\ref{Iso} provides the single element port-port isolation of the AIA. In Fig.~\ref{Iso}, P1 corresponds to port-1 of the lower band antenna antenna radiating across 6 - 12\, \si{GHz}. P2 corresponds to the midband antenna operating between 12 - 18\,\si{GHz}, and P3 represents the ports of the patch antenna designed to operate between 18 - 24\,\si{GHz}. As expected, the isolation between monopole (P2) and the patches(P3) is the maximum due to their increased physical separation, in comparison to the placement of the dipoles. Nevertheless, the overall isolation or mutual coupling between the bands of AiA is still better than \SI{-18}{dB}. Finally, Fig.~\ref{Rad} shows the radiation pattern at the center frequencies of each band, specifically at \SI{7.5}{GHz}, \SI{15}{GHz}, and \SI{21}{GHz}. It should be noted that the radiation pattern corresponds to a single antenna element within each band.

\section{Open Research Problems}

The upper mid-band presents an enormous
potential for cellular systems to deliver
high data rates with consistent coverage
and uniformity.  Nevertheless, significant
technical challenges remain to realize such systems.  We summarize
some open research problems indicated by the
preliminary studies in this paper.

\subsection{Channel measurements and capacity analyses} While there has been extensive channel measurements in the upper mid-band,
certain aspects need further study.  Most importantly, most of the measurement campaigns described in Section~\ref{sec:chanmeas} captured the omni-directional path loss.  Phased array systems,
similar to those now widely-used in mmWave bands, could provide insights into the spatial
structure of the channel which will be necessary for modeling MIMO and beamforming.
The dynamics of blockage, across the 
band, will also need to be investigated, given
how challenging blockage has been for mmWave systems.

Also, our study in Section~\ref{sec:multibandcap} suggested the possibility of significant gains with 
wideband adaptive systems operating
across the upper mid-band.  However,
the study was limited to a single
urban area with only outdoor users.
To analyze how general these results are,
statistical models and more data, validated
through measurements, will be needed.
Current statistical models, such as those
used by 3GPP \cite{3GPP38901}, 
will also need to be extended.  
Large-scale statistical dependencies between multiple bands are not well modeled,
and data-driven techniques, such as those recently proposed in \cite{hu2022multi} may be valuable. 

\subsection{Interference with incumbents}
Similarly, our analysis in Section~\ref{sec:sat_int} suggested that
high density LEO constellations
can be susceptible to interference from
terrestrial cellular services, but
spatial nulling may be able to mitigate
these effects.  Further research, however,
is required.  Interference nulling will require tracking, including tracking of NLOS components, such as ground reflections, which can be significant when one wants to suppress interference by more than \SI{20}{dB}.
Protocols that can also selectively avoid the time and frequency of transmissions that cannot
be mitigated will also be needed.  Further work will be needed for other constellations
as well as interference with radio astronomy and other passive sources, which we have only briefly mentioned.

\subsection{Antennas and circuits}
The design presented in Section~\ref{sec:antenna} indicates the possibility of a compact multiband antenna structure that can cover the entire wideband
with appropriate RF switching.  One limitation
in the current design is that the elements
are assumed to be probe fed and further work
will be needed to build
microstrip fed structures and packaging
to realize such antennas in practical devices.
In addition, tightly coupled arrays require
signal processing to account for the mutual coupling at lower bands.  In addition,
we have not addressed the design of the
RF circuits
and switches that will need to operate
across a wideband with a large number of
antenna elements.

\subsection{Security and resiliency}
Due to space considerations, this article has not
touched on the vital topic of security, a key area mentioned in the FCC study
\cite{fcc20236Gworkinggroup}.
Spectrally agile systems in the upper mid-band
could provide new resiliency to hostile attacks
by sensing signals and frequency hopping.
There is broad literature on such systems,
but little has been researched specifically
in the context of cellular multi-band
systems.  Furthermore, we have seen that
satellite signals can be significantly impacted
by low-power random terrestrial signals at
these frequencies.  This fact suggests that an
adversarial attacker could significantly disrupt vital LEO satellite services.
Methods to detect and mitigate such attacks,
possibly leveraging terrestrial measurements,
will be an area of paramount importance.

\section{Conclusion}

We have provided a detailed assessment of 
both the potential benefits and challenges that may arise for cellular use in the upper-mid band.  This analysis yields valuable insights:

{
\emph{Capacity gains:} The potential capacity gains of a wideband FR3 was assessed in an urban scenario. As expected, we showed that the overall system capacity was maximized when lower frequencies are dedicated to cell edge users, due to the more favorable propagation features. Similarly, such frequencies are indispensable for indoor UEs, owing to the substantial penetration losses in higher frequencies introduced by materials such as walls and glass. On the other hand, the key benefits of the higher frequencies stem from (1) the availability of high bandwidths and (2) the intrinsic directionality, and hence interference isolation, which translate in higher data rates.  These results demonstrate
the value of systems that can dynamically select frequency across the upper mid-band.}

\emph{Coexistence:} Then, we analyze the coexistence between cellular services and satellite incumbents, and conclude that interference may lead to substantial degradation in the performance of satellite networks.
For this reason, we propose an interference nulling scheme that enables terrestrial networks to significantly reduce that interference.

{\emph{Antenna design:} A compact wide-band
antenna array is presented based on a low-profile aperture-in-aperture (AiA) realization and three classes of antennas (coupled dipole array, circular monopole, and UWB patches).  The presented aperture provides contiguous coverage across the entire FR3 bands.}
\section*{Acknowledgment}

The authors would like to thank Muhammad Mubasshir Hossain for his assistance with the antenna design and the CAD tool.
\bibliographystyle{IEEEtran}
\bibliography{references_wireless,bibarjuna,bib_greg,bibSatheesh}
\begin{IEEEbiography}[
{\includegraphics[width=1in,height=1.25in,clip,keepaspectratio]{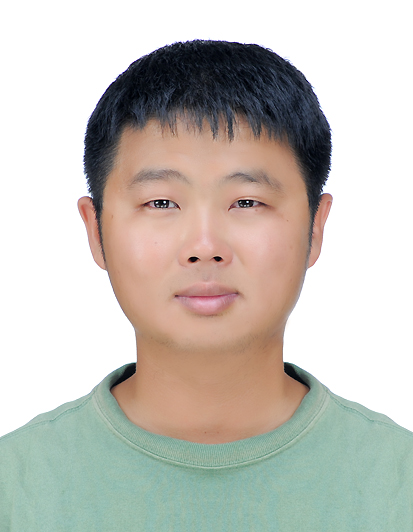}}]{Seongjoon Kang}\,\  (Graduate Student Member, IEEE) 
 received a B.S. and M.S. degrees in electrical and computer engineering from the Seoul National University, in 2017 and 2019. He is currently pursuing the Ph.D. degree with the Tandon School of Engineering, New York University, under the supervision of Professor Sundeep Rangan. His research interests include UAV and satellite communication,  MIMO, and upper mid-band and millimeter wave spectrum management.
\end{IEEEbiography}

\begin{IEEEbiography}
[{\includegraphics[width=1in,height=1.25in,clip,keepaspectratio]{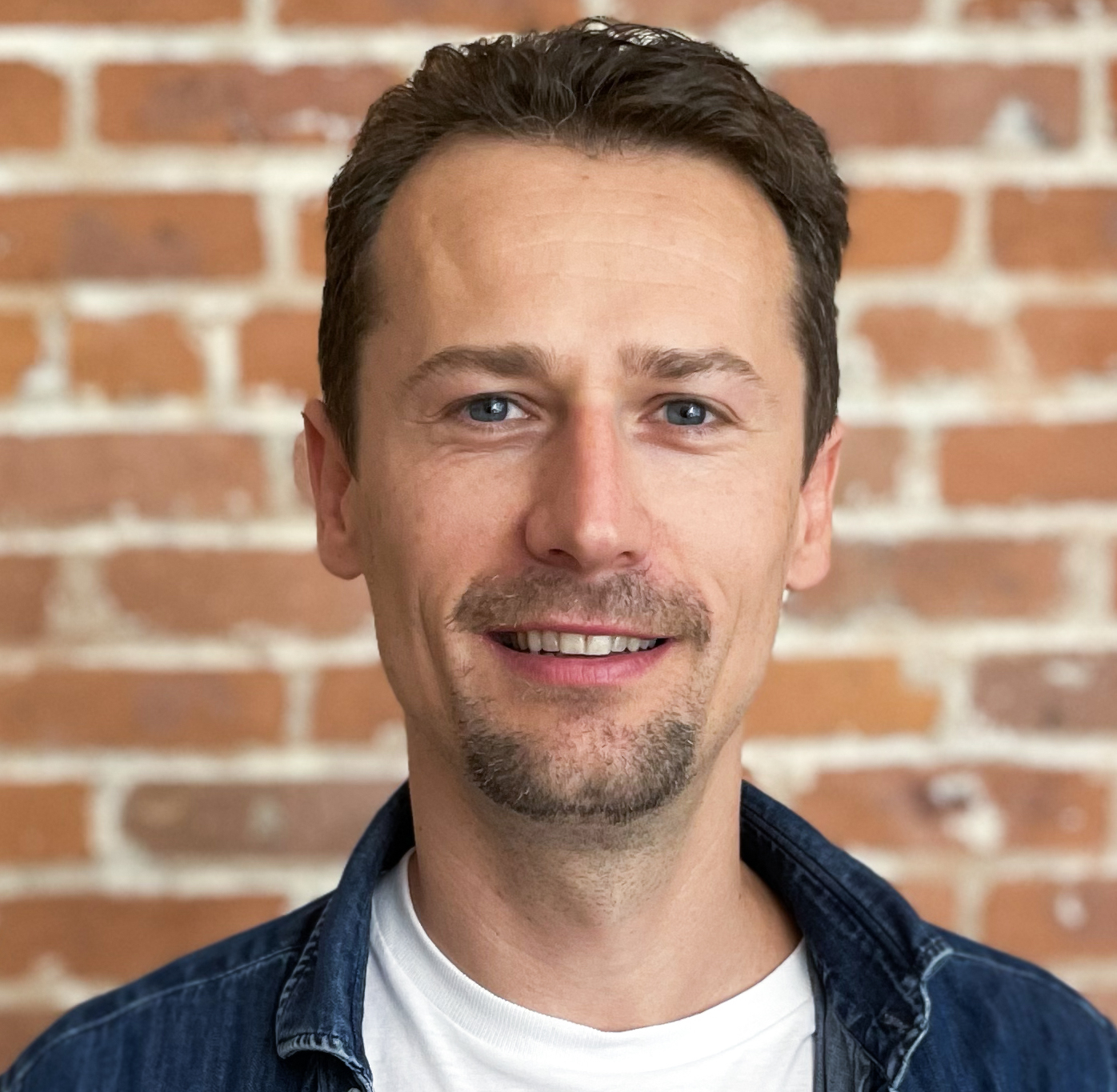}}]{Marco Mezzavilla} \,\ (Senior Member, IEEE) received the B.Sc., M.Sc., and Ph.D. at the University of Padua, Italy, in Electrical Engineering. He held visiting research positions at the NEC Network Laboratories in Heidelberg, at the Centre Tecnològic Telecomunicacions Catalunya in Barcelona, and at Qualcomm Research in San Diego. He joined New York University in 2014, where he is currently a Research Faculty. He leads several research projects that focus on upper mid-band, mmWave, and sub-THz radio access technologies for next generation wireless systems. His research interests include communication protocols, wireless prototyping, cybersecurity, and robotics.
\end{IEEEbiography}

\begin{IEEEbiography}[{\includegraphics[width=1in,height=1.25in,clip,keepaspectratio]{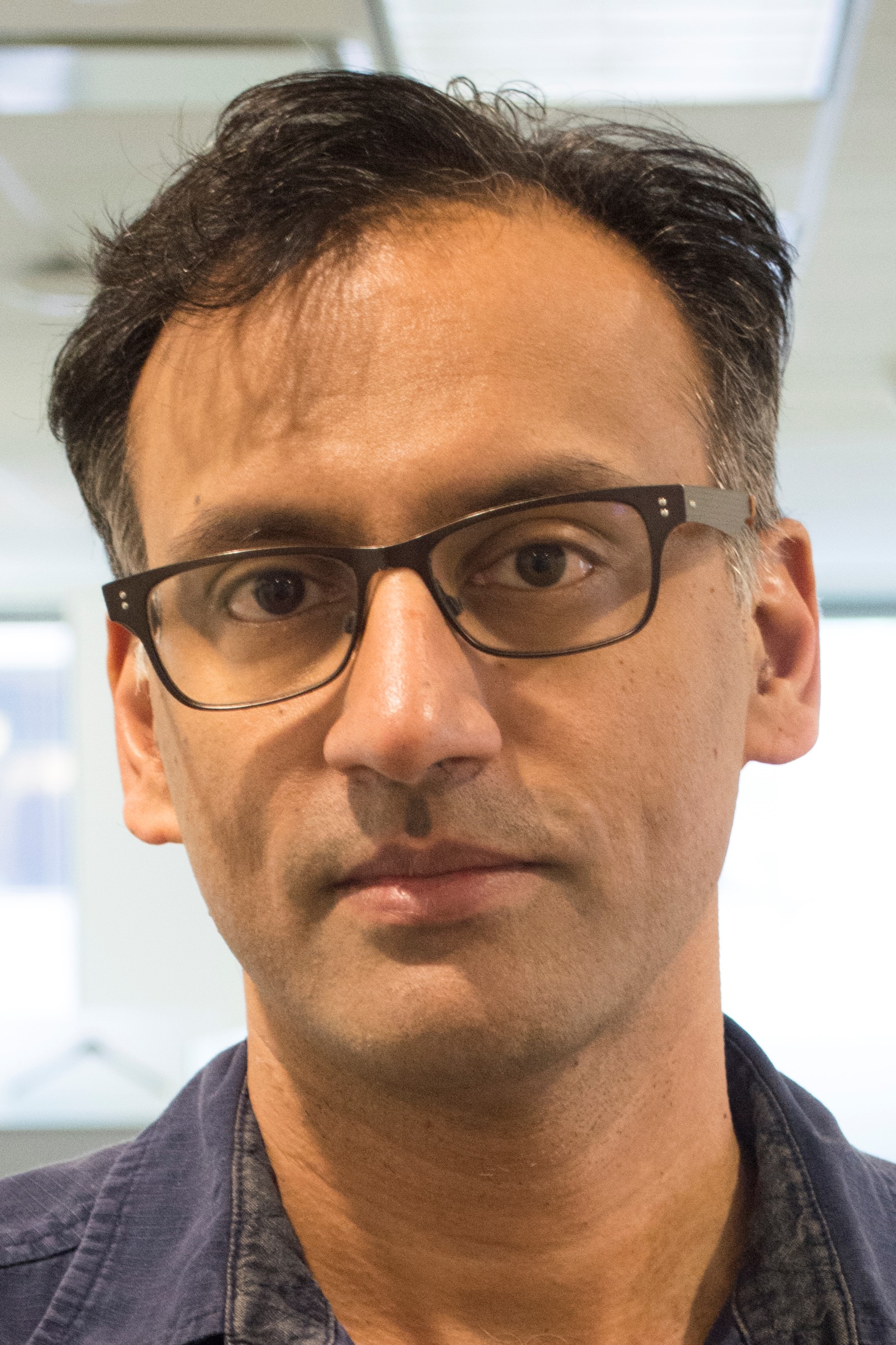}}]{Sundeep Rangan}\,\  (Fellow, IEEE)  received the B.A.Sc. at the University of Waterloo, Canada and the M.Sc. and Ph.D. at the University of California, Berkeley, all in Electrical Engineering. He has held postdoctoral appointments at the University of Michigan, Ann Arbor and Bell Labs.  In 2000, he co-founded (with four others) Flarion Technologies, a spin-off of Bell Labs, that developed Flash OFDM, the first cellular OFDM data system and pre-cursor to 4G cellular systems including LTE and WiMAX. In 2006, Flarion was acquired by Qualcomm Technologies.  Dr. Rangan was a Senior Director of Engineering at Qualcomm involved in OFDM infrastructure products. He joined NYU Tandon (formerly NYU Polytechnic) in 2010 where he is currently a Professor of Electrical and Computer Engineering.  He is a Fellow of the IEEE and the Associate Director of NYU WIRELESS, an industry-academic research center on next-generation wireless systems.
\end{IEEEbiography}

\begin{IEEEbiography}
[{\includegraphics[width=1in,height=1.25in,clip,keepaspectratio]{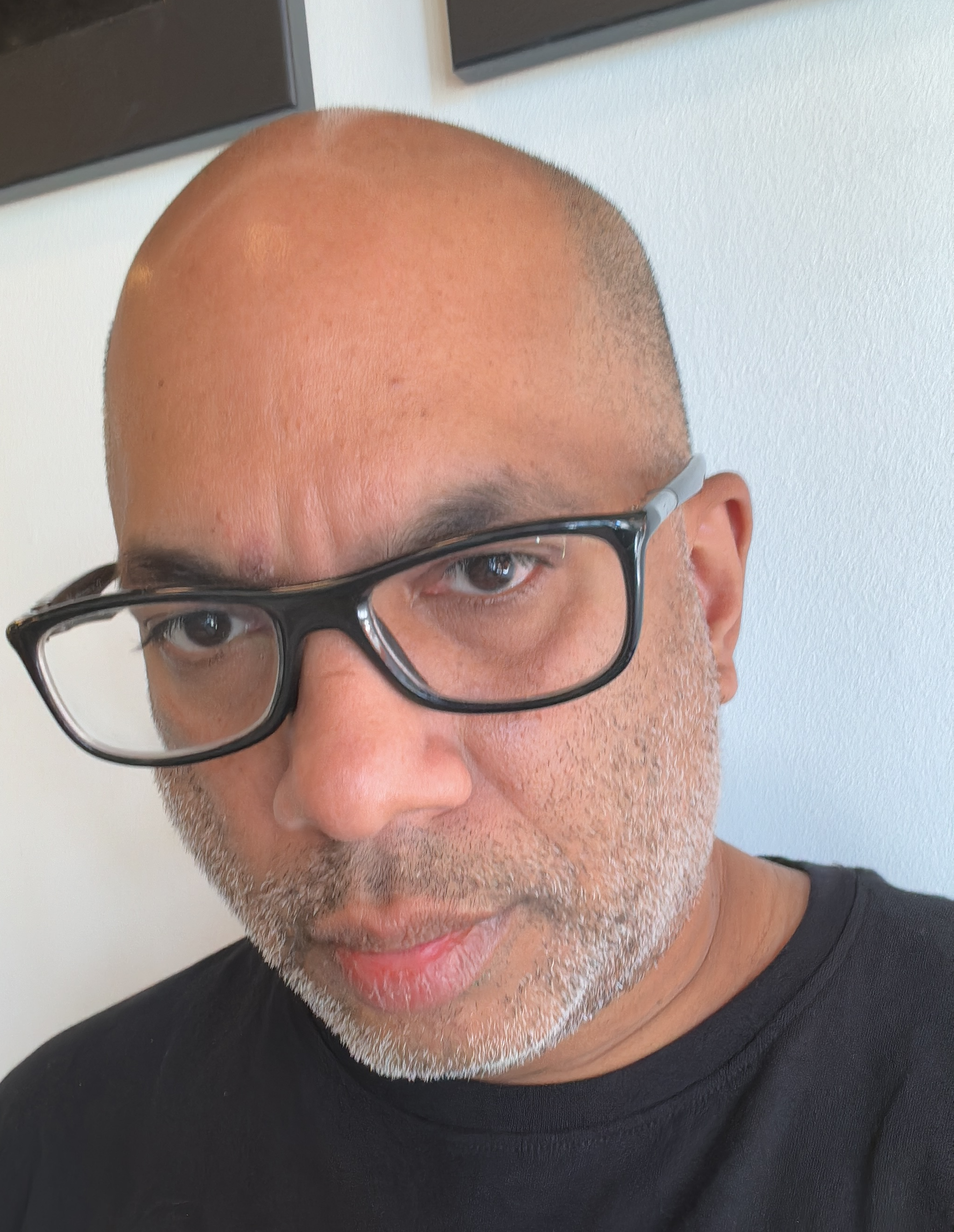}}]{Arjuna Madanayake} \,\ (Member, IEEE) is an Associate Professor of Electrical and Computer Engineering at Florida International University (FIU), Miami, FL. He directs the
RF, Analog and Digital (RAND) lab where he advises about 13 PhD students on various topics supported by NSF SWIFT, NSF FuSE, NSF MRI, ONR, NIH, Digital Locations, NTIA, Lockheed Martin, CIA Labs, NSF ICORPS, NSF IUCRC and NSF SpectrumX. His research areas span the intersection of RF and analog CMOS circuits, digital ASIC and RF-SoC/FPGAs, mixed-signal and microwave/mm-wave system design. He is interested in arrays, signal processing and computer architecture. Dr. Madanayake completed PhD and M.Sc. degrees, both in electrical engineering, from the University of Calgary, Canada, and the B.Sc. in Electronic and Telecommunication Engineering from the University of Moratuwa, Sri Lanka. He is a founding member of the IEEE Circuits and Systems Education and Outreach (CASEO) Technical Committee, and a member of the IEEE Technical Committee on Digital Signal Processing. He is the founder of Arcane AI and Wireless which is a startup focused on commercialization activities for research conducted at RAND Lab. 
\end{IEEEbiography}

\begin{IEEEbiography}
[{\includegraphics[width=1in,height=1.25in,clip,keepaspectratio,angle=0]{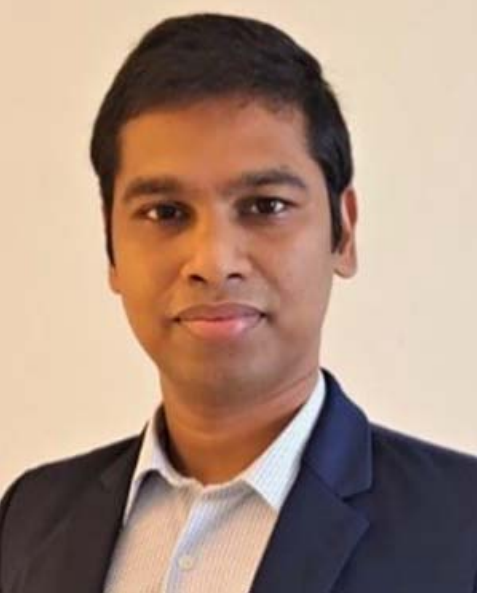}}]
    {Satheesh Bojja Venkatakrishnan} \,\ (Senior Member, IEEE)
 was born in Tiruchirappalli, India, in 1987. He received the bachelor’s degree in electronics and communication engineering from the National Institute of Technology, Tiruchirappalli, India, in 2009, and the M.S. and Ph.D.
degrees in electrical engineering from the Ohio State University, Columbus, OH, USA, in 2017.
He was a Scientist for DRDO, India, from 2009 to 2013, working on the development and implementation of active electronic steerable antennas. He is currently an Assistant Professor of electrical and computer engineering with Florida International University, Miami, FL, USA. His current research interests include RF system design for secure wideband communications, data sensing and imaging, interference mitigation techniques, and RFSoC based
simultaneous transmit and receive system (STAR) to improve the spectral
efficiency. In parallel, he has been working on developing RF sensors and
circuits including fully passive neural implants and multi-modal patch sensors
for bio-medical applications. Dr. Bojja Venkatakrishnan was the recipient
of numerous awards and recognitions including the IEEE Electromagnetic
Theory Symposium (EMTS-2019) Young Scientist Award, and the Best Paper
Award in the International Union of Radio Science General Assembly and
Scientific Symposium (URSI-GASS) held in Montreal, Canada in August
2017. He is a member of Phi Kappa Phi, and also an associate member of
USNC–URSI.
\end{IEEEbiography}

\begin{IEEEbiography}
[{\includegraphics[width=1in,height=1.25in,clip,keepaspectratio,angle=0]{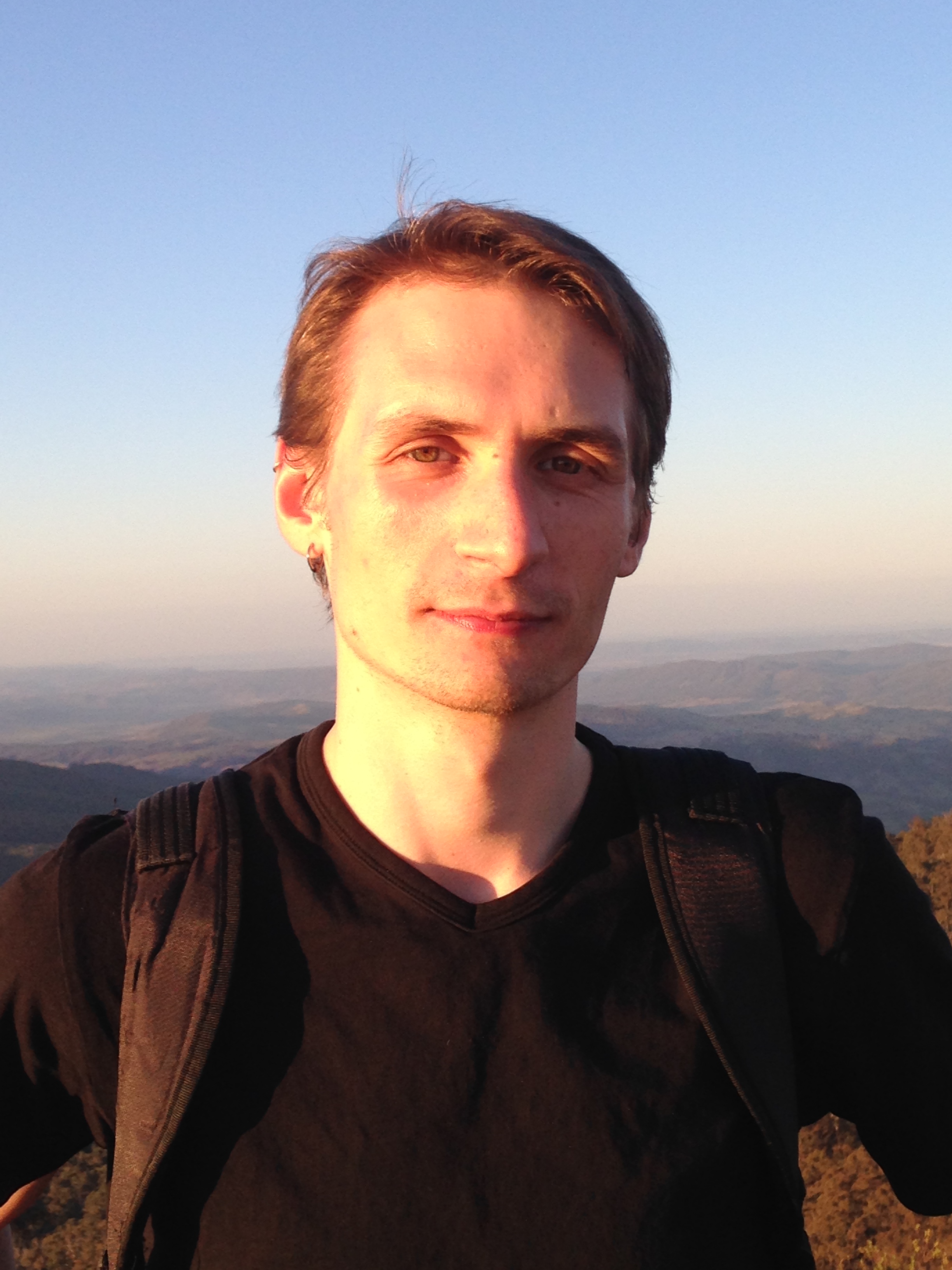}}]
{Gr{\'e}gory Hellbourg}\,\,received the engineering and M.Sc. degree in signal and image processing and the Ph.D. degree in signal processing and automation from the University of Orléans, France, in 2010 and 2014, respectively. He is a Staff Scientist with the Cahill Center for Astronomy and Astrophysics, California Institute of Technology, California, USA, and a Spectrum Manager of the Owens Valley Radio Observatory, California. He will oversee the main decisions and act as a Project Manager. His expertise lies in signal processing, radio interference detection and mitigation, and system design for radio astronomy.
\end{IEEEbiography}

\begin{IEEEbiography}[{\includegraphics[width=1in,height=1.25in,clip,keepaspectratio]{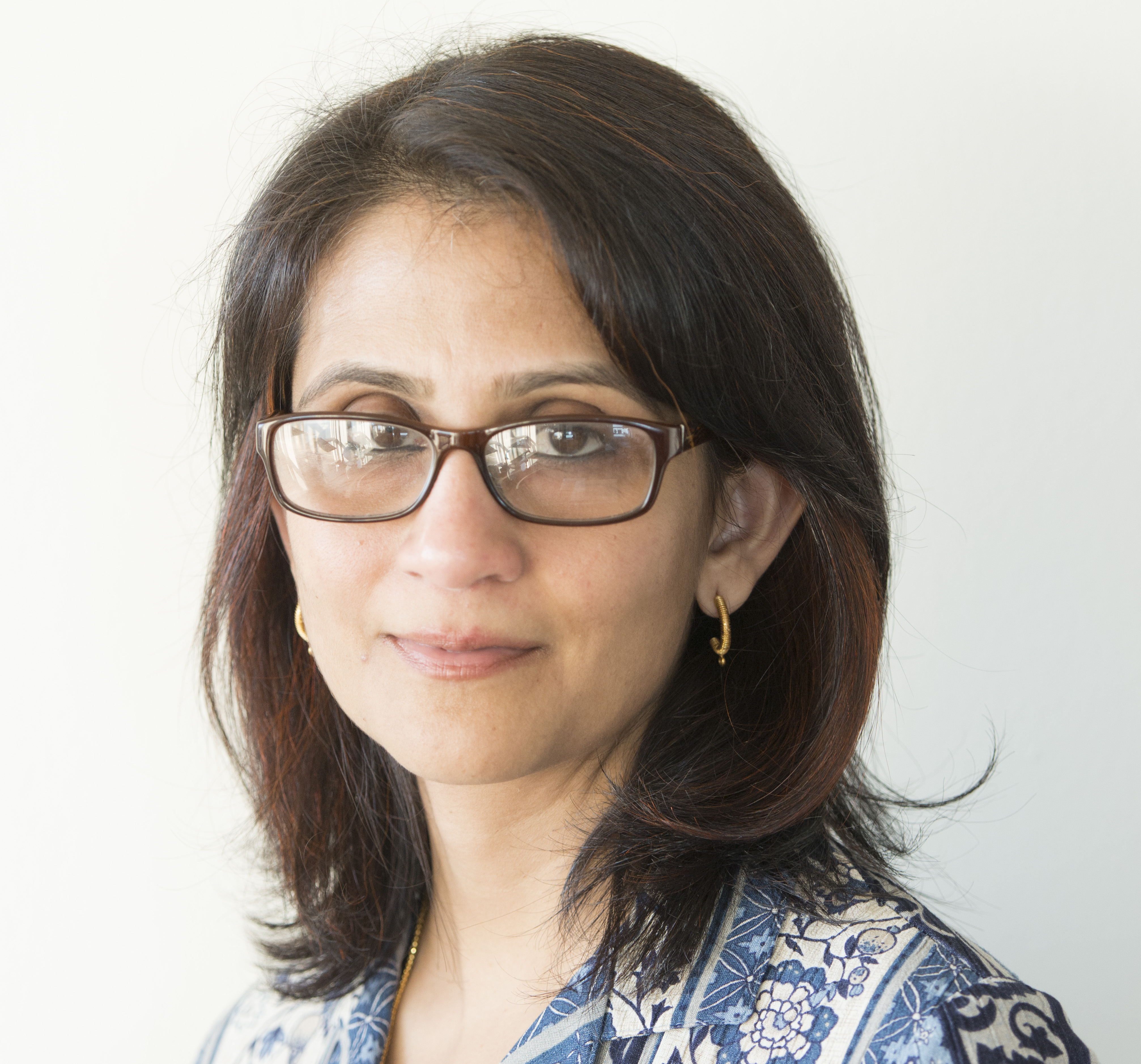}}]{Monisha Ghosh}\,\  (Fellow, IEEE)
 is a Professor of Electrical Engineering at the University of Notre Dame and a member of the Notre Dame Wireless Institute. She is also the Policy Outreach Director for SpectrumX (https://www.spectrumx.org/ ),the first NSF Center for Spectrum Innovation. Her research interests are in the development of next generation wireless systems: cellular, Wi-Fi, and loT, with an emphasis on spectrum sharing, coexistence, and applications of machine learning to improve network performance. Prior to joining the University of Notre Dame in 2022, she was the chief technology officer at the Federal Communications Commission, a Program Director at the National Science Foundation, Research Professor at the University of Chicago. She also spent 24 years in industry research at Bell Labs, Philips Research and Interdigital working on a wide variety of wireless systems: HDTV, Wi-Fi, TV white spaces, and cellular. She obtained her B.Tech from IIT Kharagpur, in 1986, and Ph.D. from USC, in 1991.
\end{IEEEbiography}

\begin{IEEEbiography}
[{\includegraphics[width=1in,height=1.25in,clip,keepaspectratio]{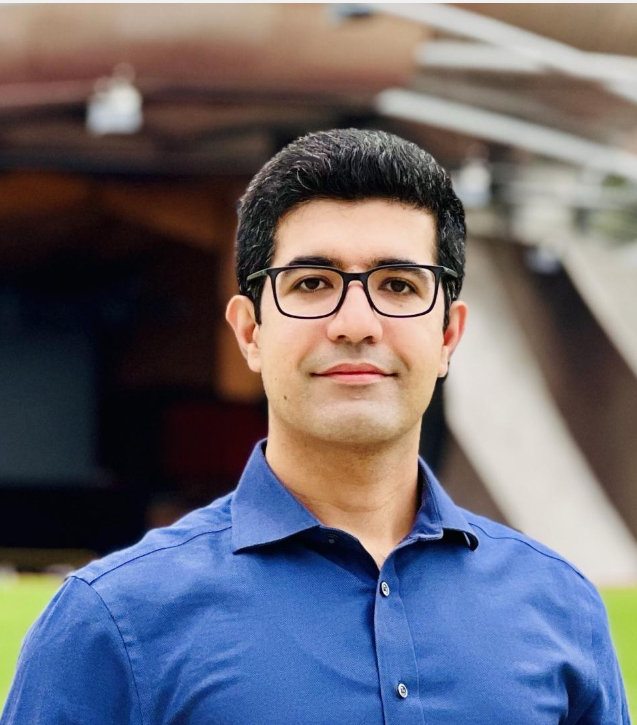}}]
{Hamed Rahmani}\,\ (Member, IEEE)
 is an Assistant Professor of Electrical and Computer Engineering at New York University (NYU). He received a Ph.D. degree from the University of California Los Angeles (UCLA), an M.Sc. degree from Rice University, Houston, TX, and a B.Sc. degree from Sharif University of Technology, Tehran, Iran, all in Electrical and Computer Engineering. Before joining NYU, he held multiple industry and research positions. As a research scientist, he worked with IBM T. J. Research Center in Yorktown Heights on high-speed electrical/optical interconnects. He was an Adjunct Professor at Columbia University in New York, NY, and a visiting lecturer at Princeton University, where he offered graduate-level courses in analog and RF circuit design. He was also a senior RFIC design engineer at Qualcomm Inc., where he focused on advanced 5G transmitters for cellular applications and RF front-end designs. Dr. Rahmani is the recipient of several prestigious awards and fellowships, including the IEEE MTT-S Graduate Fellowship for medical applications and the Texas Instruments Distinguished Fellowship.
\end{IEEEbiography}

\begin{IEEEbiography}
[{\includegraphics[width=1in,height=1.25in,clip,keepaspectratio]{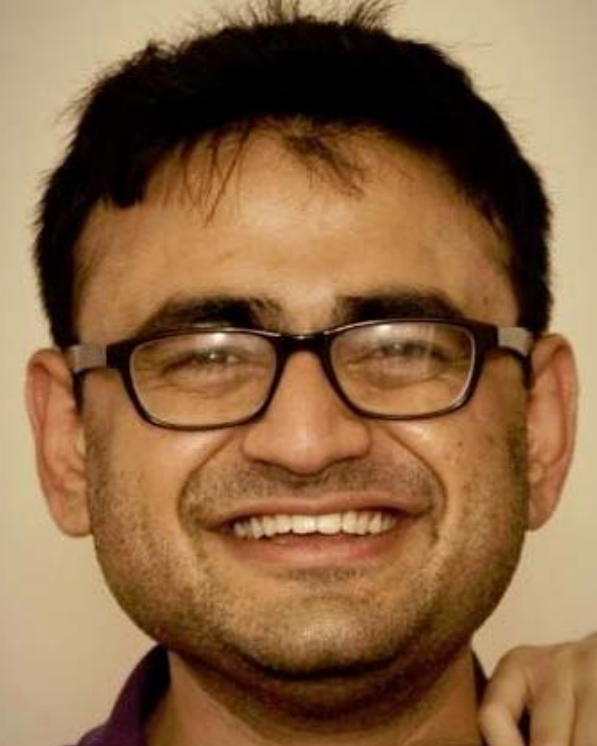}}]
{Aditya Dhananjay}\,\,received the Ph.D. degree from the Courant Institute of Mathematical Sciences, New York University (NYU), New York, NY, USA. He was involved in mesh radio routing and resource allocation protocols, data communication over cellular voice channels, low-cost wireless rural connectivity, OFDM equalization, and phase noise mitigation in mm-wave networks. He currently holds a post-doctoral position with NYU. He has developed and supervised much of the mm-wave experimental work at the center. He has authored several refereed articles (including at SIGCOMM and MobiCom). He holds one patent and two provisional patents in the millimeter-wave space.
\end{IEEEbiography}

\end{document}